\documentclass[aps,twocolumn,nofootinbib,showkeys,notitlepage,superscriptaddress]
{revtex4-2}
\usepackage{graphicx}
\usepackage{dcolumn}
\usepackage{bm}
\usepackage{amsmath}
\usepackage{float}
\usepackage{multirow}
\usepackage{slashed}
\usepackage{xcolor}
\usepackage{physics}
\usepackage{gensymb}
\usepackage{mathtools}
\usepackage{braket}
\usepackage{lipsum}
\usepackage{amsfonts}
\usepackage{amssymb}
\usepackage{epsfig}
\usepackage{ulem,tensor}
\usepackage{tabularx}

\usepackage[colorlinks=true,pdfstartview=FitV,bookmarks=true,bookmarksnumbered=true,
breaklinks]{hyperref}
\usepackage{color}

\definecolor{blue}{rgb}{0.0, 0.0, 1.0}
\definecolor{red}{rgb}{1.0, 0.0, 0.0}
\definecolor{royalblue}{rgb}{0.0, 0.14, 0.4}
\hypersetup{linkcolor=red, citecolor=blue, urlcolor=blue}

\def\orcid#1{\kern .08em\href{https://orcid.org/#1}
{\includegraphics[keepaspectratio,width=0.7em]{ORCID_iD.png}}}
\usepackage{calligra}
\DeclareMathAlphabet{\mathcalligra}{T1}{calligra}{m}{n}

\def\la{\langle}
\def\ra{\rangle}

\def\be{\begin{equation}}
\def\ee{\end{equation}}
\def\bea{\begin{eqnarray}}
\def\eea{\end{eqnarray}}

\def\la{\langle}
\def\ra{\rangle}

\def\be{\begin{equation}}
\def\ee{\end{equation}}
\def\bea{\begin{eqnarray}}
\def\eea{\end{eqnarray}}

\renewcommand\sout{\bgroup \color{red} \ULdepth=-.5ex \ULset}

\begin{document}
\title{\Large Exclusive $J/\psi$ photo-production on nuclei}
\author{Sang-Ho Kim}
\email{shkimphy@gmail.com}
\affiliation{Department of Physics and Origin of Matter and Evolution of Galaxies
(OMEG) Institute, Soongsil University, Seoul 06978, Korea}
\author{T.-S. H. Lee}
\email{tshlee@anl.gov}
\affiliation{Physics Division, Argonne National Laboratory, Argonne, Illinois 60439, USA}
\author{R.B. Wiringa}
\email{tshlee@anl.gov}
\affiliation{Physics Division, Argonne National Laboratory, Argonne, Illinois 60439, USA}

\date{\today}
\begin{abstract}
Motivated by the recent experimental developments, the Pom-CQM model of the $\gamma
+ N \to J/\psi + N$ reaction of Lee~\textit{et al.}
\href{https://doi.org/10.1140/epja/s10050-022-00901-9}{[Eur. Phys. J. A. \textbf{58},
252 (2022)]}
and Sakinah~\textit{et al.}
\href{https://journals.aps.org/prc/abstract/10.1103/PhysRevC.109.065204}{
[Phys. Rev. C. \textbf{109}, 065204 (2024)]} has been applied to predict the exclusive
$J/\psi$ photo-production on nuclei ($A$).
Within the multiple scattering theory, the calculations have been performed by
including the impulse amplitude $T^{\rm IMP}_{J/\psi A,\gamma A}$ and the
$J/\psi$-nucleus final state interaction (FSI) amplitude $T^{\rm FSI}_{J/\psi A,\gamma A}$. 
For the deuteron target, $T^{\rm IMP}_{J/\psi d,\gamma d}$ is calculated exactly using
the wave function generated from the realistic nucleon-nucleon potentials.
It is found that, near the threshold region, the $J/\psi$ photo-production cross
sections depend sensitively on the $d$-state of the deuteron wave function.
The FSI amplitude $T^{\rm FSI}_{J/\psi A,\gamma A}$ is calculated using the first-order
optical potential constructed from the $J/\psi$-$N$ scattering amplitude generated
from the employed Pom-CQM model.
It turns out that the FSI has significant effects in the large momentum-transfer
region.
By using the conventional fixed scatter approximation (FSA) and the nuclear form
factors from the variational Monte-Carlo (VMC) calculations of
Lonardoni~\textit{et al.}
\href{https://journals.aps.org/prc/abstract/10.1103/PhysRevC.96.024326}{
[Phys. Rev.C. \textbf{96}, 024326 (2017)]}, the
cross sections of the $J/\psi$ photo-production on ${^4\rm He}$, ${^{16}\rm O}$, and
${^{40}\rm Ca}$ are also predicted for future experimental investigations at JLab and
EIC.
\end{abstract}


\maketitle

\section{Introduction} \label{Section:I}

It is well recognized~\cite{Lee:2022ymp} that the study of $J/\psi$ photo-production on
the nucleon can be used to explore the role of gluons in determining the properties of
the nucleon and hadron-hadron interactions.
The experimental data of $\gamma + p \to J/\psi + p$ at photon energies near the
threshold region ($8.90 \leqslant E_\gamma \leqslant 10.90$ GeV) have recently been
reported by the Jefferson Laboratory (JLab)~\cite{GlueX:2019mkq,GlueX:2023pev,
Duran:2022xag} and have been analyzed within several theoretical
models~\cite{Brodsky:2000zc,Mamo:2019mka,Du:2020bqj,Lee:2020iuo,Guo:2021ibg,
JPAC:2023qgg} and also recently in Refs.~\cite{Sakinah:2024cza,Tang:2024pky}. 
With the developing experimental and theoretical efforts, the data of $J/\psi$
photo-production on nuclei will be also available at 
JLab~\cite{Arrington:2021alx,Accardi:2023chb,Liu:2023htv,Tyson:2024wbx,Pybus:2024ifi}
in the near future, and eventually at Electron-Ion Collider
(EIC)~\cite{Accardi:2012qut,Abir:2023fpo,Wang:2023thy}.

The EIC has the capability to measure the diffractive and exclusive processes with a
variety of ion beams and will provide the first 3-dimensional images of sea
quarks and gluons in a fast-moving nucleus with sub-femtometer resolution.
For example, the EIC can obtain the spatial distribution of gluons in a nucleus by
measuring the coherent diffractive production of $J/\psi$ in electron-nucleus
scattering, as in the case of electron-proton scattering.
Coherent exclusive processes, in which the nucleus stays intact, can give us valuable
insights into understanding collective dynamics such as shadowing, anti-shadowing
or the EMC effect~\cite{Accardi:2012qut}.

The purpose of this work is to help facilitate these efforts by applying the
Pom (Pomeron)-CQM (constituent quark model) model of the $\gamma + N \to J/\psi + N$
reaction developed in Refs.~\cite{Lee:2022ymp,Sakinah:2024cza} to predict the cross
sections of the exclusive $J/\psi$ photo-production on nuclei ($A$) and to investigate
how these cross sections are related to the nuclear structure, as a step to develop an
approach to investigate the role of gluons in nuclei.
We will perform the calculations within the well developed multiple scattering
theory~\cite{Watson:1960,Feshbach:1992,Kerman:1959fr,Landau:1973iv,Lee:1974zr} and
will present the predicted cross sections of the exclusive $J/\psi$ photo-production
on the deuteron ($d$), ${^4\rm He}$, ${^{16}\rm O}$, and ${^{40}\rm Ca}$ targets.
For $A > 2$ nuclei, we extend our previous work on $\gamma + {^4\rm He} \to \phi +
{^4\rm He}$~\cite{Kim:2021adl} where the LEPS (Laser Electron Photon Experiment at
SPring-8) data~\cite{LEPS:2017nqz} are fairly accounted for.
Previous theoretical studies on $\gamma + A \to J/\psi + A$ are very
rare~\cite{Hatta:2019ocp,He:2024lry}.
We are also motivated by Ref.~\cite{Hatta:2019ocp} to explore how the nuclear short
range correlations can be probed by $J/\psi$ photo-production reactions.

This paper is organized as follows.
In Sec.~\ref{Section:II}, we briefly review the Pom-CQM model constructed in
Refs.~\cite{Lee:2022ymp,Sakinah:2024cza}.
The formulation of $J/\psi$ photo-production on nuclei is presented in
Sec.~\ref{Section:III}.
The numerical results are presented in Sec.~\ref{Section:IV} for the deuteron target
and in Sec.~\ref{Section:V} for the ${^4\rm He}$, ${^{16}\rm O}$, and
${^{40}\rm Ca}$ targets.
In Sec.~\ref{Section:VI}, we summarize the present work and discuss possible
improvements which are needed to investigate the roles of gluons in nuclei using the
data from  future experiments.
In the Appendix, we provide  the explanation for the Pom-CQM model which is used for
the calculations in this work and is needed to develop the formulation of the $J/\psi$
photo-production reactions on nuclei.

\section{Pom-CQM model} \label{Section:II}

With the normalization~\cite{Goldberger:1975} $\la{\bf k}|{\bf k}'\ra =
\delta({\bf k}-{\bf k}')$ for plane wave state $|{\bf k}\ra$ and
$\la\phi_{\alpha}|\phi_\beta\ra=\delta_{\alpha,\beta}$ for bound state $|\phi_\alpha\ra$,
the differential cross section of the $J/\psi$ (also denoted as $V$) photo-production
reaction, in  the center of mass (c.m.) frame,
$\gamma({\bf q},\lambda) + N({-\bf q},m_s) \to
V({\bf k},m_V) + N(-{\bf k},m_{s'})$, is calculated from~\cite{Lee:2022ymp}
\begin{eqnarray}
&& \hskip -0.6cm \frac{d\sigma_{VN,\gamma N}}{d\Omega}
= \frac{(2\pi)^4}{|{\bf q}|^2}\frac{|{\bf k}|\, E_V({\bf k})E_N({\bf k})}{W}
\frac{|{\bf q}|^2E_N({\bf q})}{W}
\nonumber \\
&& \hskip -0.6cm \times 
\frac{1}{4} \sum_{m_{s'},m_s}\sum_{m_V,\lambda}
\left| \braket{{\bf k},m_V m_{s'}|T_{VN,\gamma N}(W)|{\bf q},\lambda m_s} \right|^2,
\label{eq:crst-gnjn}
\end{eqnarray}
where $m_a$ denotes the $z$-component of the spin of particle $a$,
$E_\alpha({\bf k})=\sqrt{{\bf k}^2+m^2_\alpha}$ is the energy of
a  particle $\alpha$ with mass $m_\alpha$,  and  $\lambda$ is
the helicity of photon $\gamma$.
The magnitudes $q=|\mathbf{q}|$ and $k=|\mathbf{k}|$ are defined by the invariant
mass $W=q+E_N(q)=E_V(k)+E_N(k)$. 

\begin{figure}[h] 
\centering
\includegraphics[width=0.7\columnwidth,angle=0]{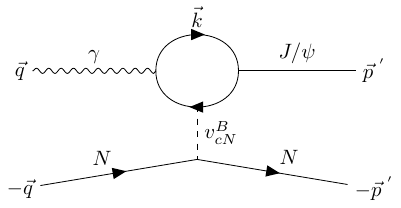}
\caption{Born term of the $J/\psi$ photo-production on the nucleon target.}
\label{fig:gnvn}
\end{figure}

\begin{figure}[h] 
\centering
\includegraphics[width=0.7\columnwidth,angle=0]{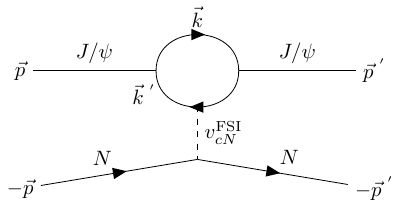}
\caption{$J/\psi N$ potential defined by the quark-nucleon potential $v_{cN}$.}
\label{fig:vnvn}
\end{figure}

Within the Pom-CQM model~\cite{Lee:2022ymp,Sakinah:2024cza}, the reaction amplitude
$T_{VN,\gamma N}(W)$ can be expressed as the sum of the dynamical scattering amplitude
$T_{VN,\gamma N}^{\rm D}(W)$ and the Pomeron-exchange amplitude $T_{VN,\gamma N}^{\rm Pom}(W)$
\begin{eqnarray}
T_{VN,\gamma N}(W) = T^{\rm D}_{VN,\gamma N}(W) + T^{\rm Pom}_{VN,\gamma N}(W).
\label{eq:t-0}
\end{eqnarray}
The amplitude $T^{\rm D}_{VN,\gamma N}(W)$, defined within a Hamiltonian 
formulation~\cite{Sato:1996gk,Matsuyama:2006rp,Kamano:2013iva},
is of the following form
\begin{eqnarray}
T^{\rm D}_{VN,\gamma N}(W) = B_{VN,\gamma N}(W) + T_{VN,\gamma N}^{\rm FSI}(W),
\label{eq:eq1-a}
\end{eqnarray}
where $B_{VN,\gamma N} (W)$ is the amplitude of the photo-production of $J/\psi$ on the
nucleon and the final state interaction (FSI) amplitude is of the following form
\begin{eqnarray}
\hskip -0.4cm
T_{VN,\gamma N}^{\rm FSI}(W) = t_{VN,VN}(W) \frac{1}{E-H_0+i\epsilon} B_{VN,\gamma N}(W).
\label{eq:eq1-b}
\end{eqnarray}
Here $H_0$ is the free Hamiltonian and $t_{VN,VN}(W)$ is the $J/\psi N \to J/\psi N$
scattering amplitude calculated from the $J/\psi N$ potential $v_{VN,VN}$ using the
following Lippmann-Schwinger equation
\begin{eqnarray}
t_{VN,VN}(W) &=& v_{VN,VN}
\nonumber \\
&+& v_{VN,VN} \frac{1}{W -H_0+i\epsilon} t_{VN,VN}(W).
\label{eq:eq1-c}
\end{eqnarray}
Within QCD, the Born term $B_{VN,\gamma N} (W)$ and $v_{VN,VN}$ are defined by the
interactions between quarks in the $J/\psi$ and nucleon due to the gluon-exchange
mechanisms.
The formulation of $B_{VN,\gamma N} (W)$ and $v_{VN,VN}$ depends on the models of the
$J/\psi$ and $N$.
The most straightforward approach is to use the CQM~\cite{Segovia:2013wma} within
which both quantities are defined by the loop-integrations of gluon-exchange
quark-quark potential $v_{qq}$ over the wave function of $J/\psi$ and $N$.
This CQM-based model was first explored in Ref.~\cite{Lee:2022ymp} and developed in
Ref.~\cite{Sakinah:2024cza} to analyze the JLab data~\cite{GlueX:2019mkq,GlueX:2023pev,
Duran:2022xag} using a simplification that the interactions are defined by the
phenomenological quark-$N$ potentials $v_{cN}(r)$.
Thus $B_{VN,\gamma N}(W)$ and $v_{VN,VN}$ are calculated from one-loop integrations
of $v_{cN}(r)$ over the $J/\psi$ wave function, as shown in Figs.~\ref{fig:gnvn}
and \ref{fig:vnvn}, respectively.

The fits to the JLab data are obtained by adjusting the parameters of the
phenomenological quark-$N$ potentials which are of the following forms
\begin{eqnarray} 
v^B_{cN}(r) = \alpha_B \left( \frac{e^{-\mu_B r}}{r}-\frac{e^{-\mu'_B r}}{r} \right),
\label{eq:qn-B}
\end{eqnarray}
for the Born term $B_{VN,\gamma N}(W)$ (Fig.~\ref{fig:gnvn}) and
\begin{eqnarray}
v^{\rm FSI}_{cN}(r) =
\alpha_{\rm FSI} \left( \frac{e^{-\mu_{\rm FSI} r}}{r}-\frac{e^{-\mu'_{\rm FSI}r}}{r} \right),
\label{eq:qn-fsi}
\end{eqnarray}
for the $J/\psi N$ potential $v_{VN,VN}$ (Fig.~\ref{fig:vnvn}).
In Eqs.~(\ref{eq:qn-B})-(\ref{eq:qn-fsi}), we have defined
{\bf  $\mu'_B= N_B \times\mu_B$} and
{\bf $\mu'_{\rm FSI}= N_{\rm FSI} \times \mu_{\rm FSI}$}.
The Pom-CQM model was explained in more detail in Refs.~\cite{Lee:2022ymp,
Sakinah:2024cza} and in Appendix.


The best fits~\cite{Sakinah:2024cza} to the JLab data of the $J/\psi$ photo-production
on the proton target are obtained from the so-called fit1 model with the parameters
\begin{eqnarray}
&&\alpha_B=-0.145;\, \alpha_{\rm FSI}=-0.1,               \nonumber \\
&&\mu_B=0.3\,{\rm GeV};\, \mu_{\rm FSI}=0.3\,{\rm GeV},   \nonumber \\
&&N_B=5;\, N_{\rm FSI}=2,                                 \nonumber\\
&&F_V(t)=1,
\label{parm:fit1} 
\end{eqnarray}
where $F_V(t) $ is the form factor of $J/\psi$ defined in Eq.~(\ref{eq:v-ff}) in
Appendix.

Figure~\ref{fig:totcrst-2y-fit1} presents the resulting total cross section of
$\gamma p \to J/\psi p$ as a function of $W$ (a) from threshold up to $W = 300$ GeV
and (b) at low energies ($4.0 \leqslant W \leqslant 5.9$ GeV).
The Pomeron exchange is in excellent agreement with the high energy data ($10
\leqslant W \leqslant 300$ GeV)~\cite{Binkley:1981kv,E687:1993hlm,ZEUS:2002wfj,
H1:1996kyo,H1:2000kis} but is not sufficent for describing the low
energy region~\cite{Camerini:1975cy,GlueX:2019mkq,GlueX:2023pev}.
The full result includes the contribution from the dynamical scattering amplitude
$T^D_{VN,\gamma N}(W)$ in Eq.~(\ref{eq:t-0}) such that the low energy data can be well
described.

\begin{figure}[ht] 
\begin{center}
\includegraphics[width=1.0\columnwidth,angle=0]{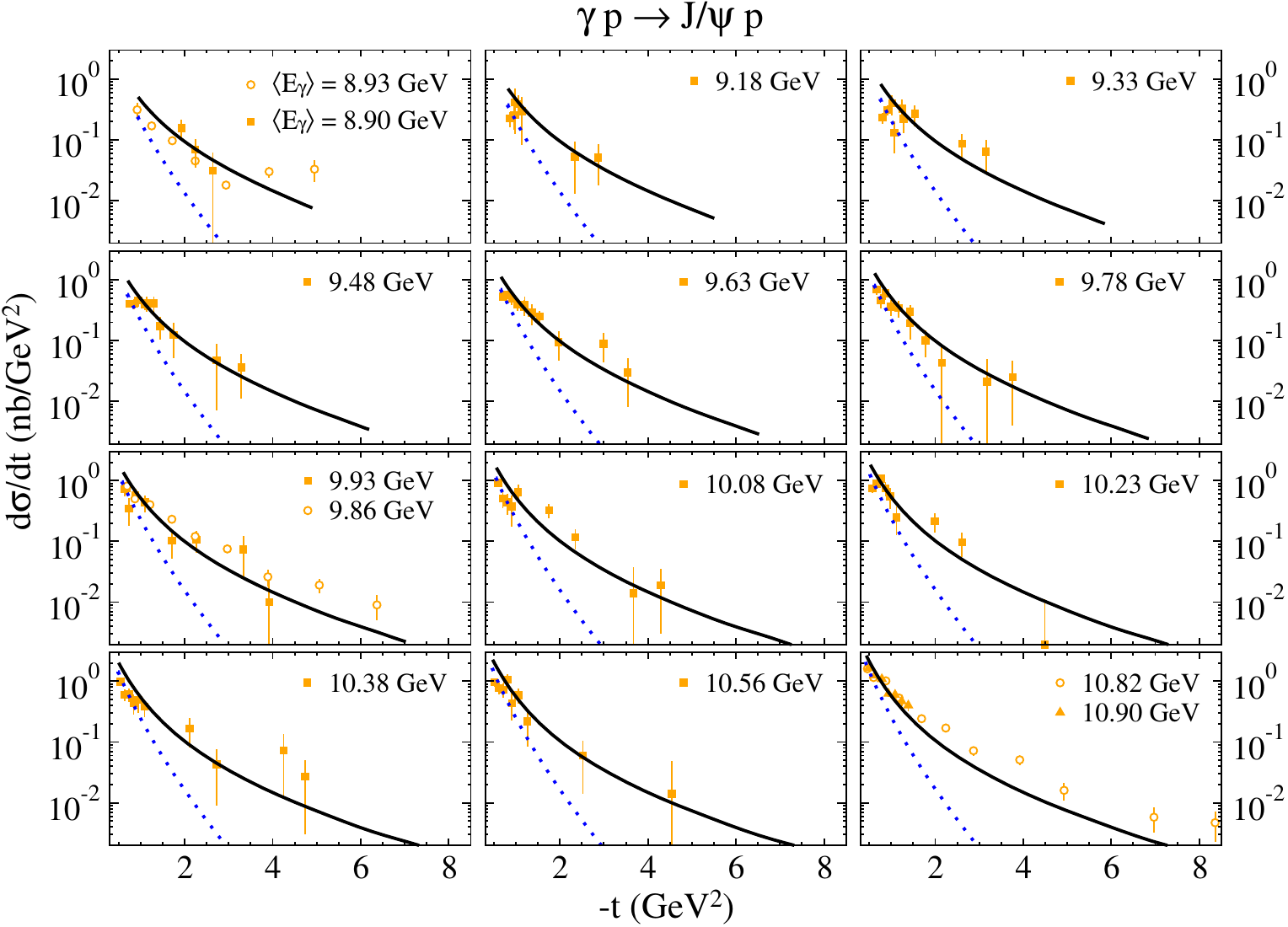}
\caption{Differential cross sections of $\gamma p \to J/\psi p$ are plotted as
functions of the momentum transfer $-t$ at different photon energies
$8.90 \leqslant E_\gamma \leqslant 10.90$ GeV.
The GleuX19~\cite{GlueX:2019mkq} (triangle), GlueX23~\cite{GlueX:2023pev} (circle), and
Hall C~\cite{Duran:2022xag} (quadrangle) data from JLab are compared with the
Pomeron-exchange (dotted curves) and full (solid curves)
contributions~\cite{Sakinah:2024cza}.}
\label{fig:dsdt-2y-fit1}
\end{center}
\end{figure}

In Fig.~\ref{fig:dsdt-2y-fit1}, the differential cross sections of
$\gamma p \to J/\psi p$ are displayed as functions of $-t$ at different photon
energies $8.90 \leqslant E_\gamma \leqslant 10.90$ GeV.
The shapes of the Pomeron exchange are too steep to reproduce the available data.
The full results agree well with the JLab data~\cite{GlueX:2019mkq,GlueX:2023pev,
Duran:2022xag}.
Thus the fit1 model in Ref.~\cite{Sakinah:2024cza} is suitable for developing an
approach to investigate $J/\psi$ photo-production on nuclei.


\begin{figure}[t] 
\includegraphics[width=0.85\columnwidth,angle=0]{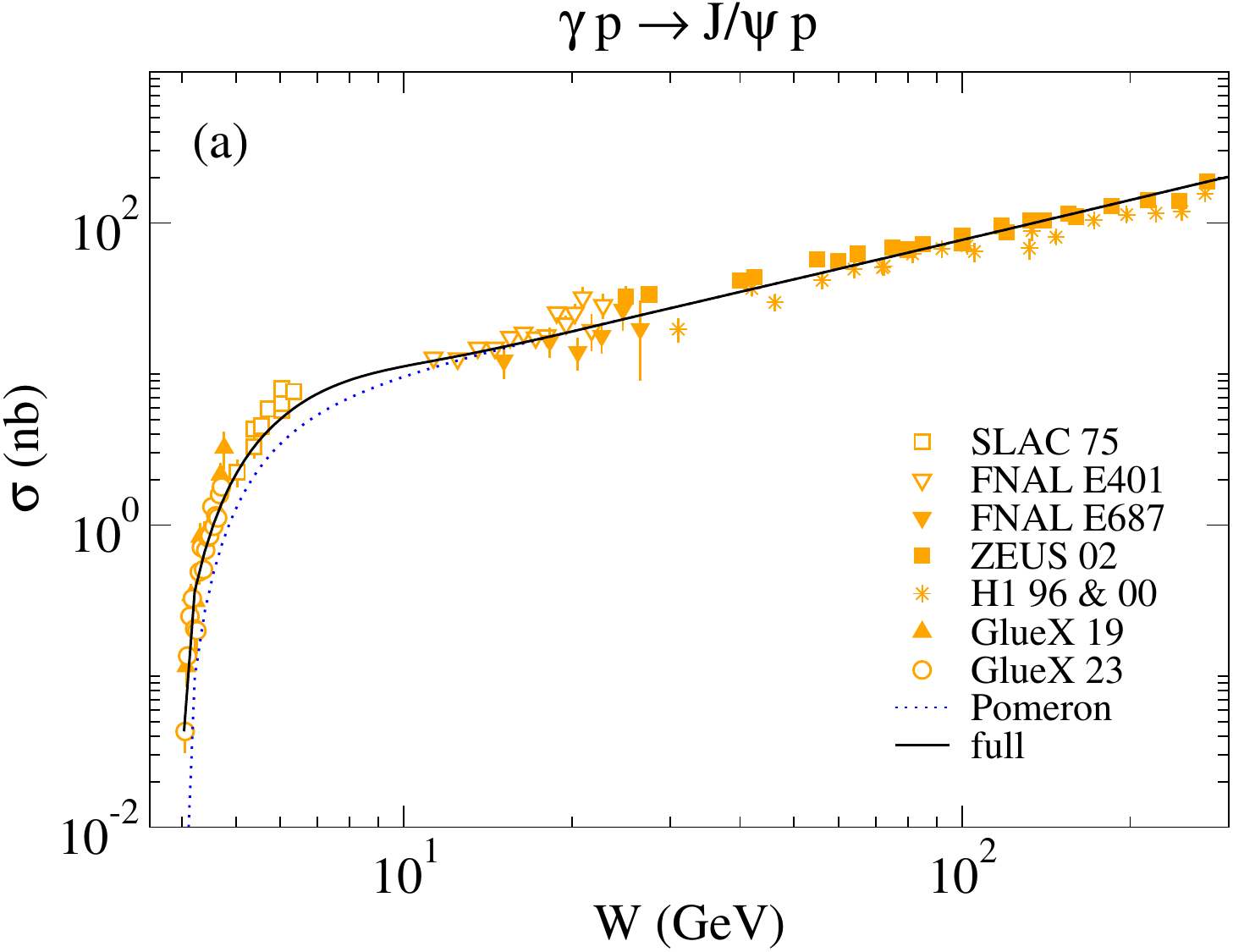} \\
\vspace{0.5em}
\includegraphics[width=0.85\columnwidth,angle=0]{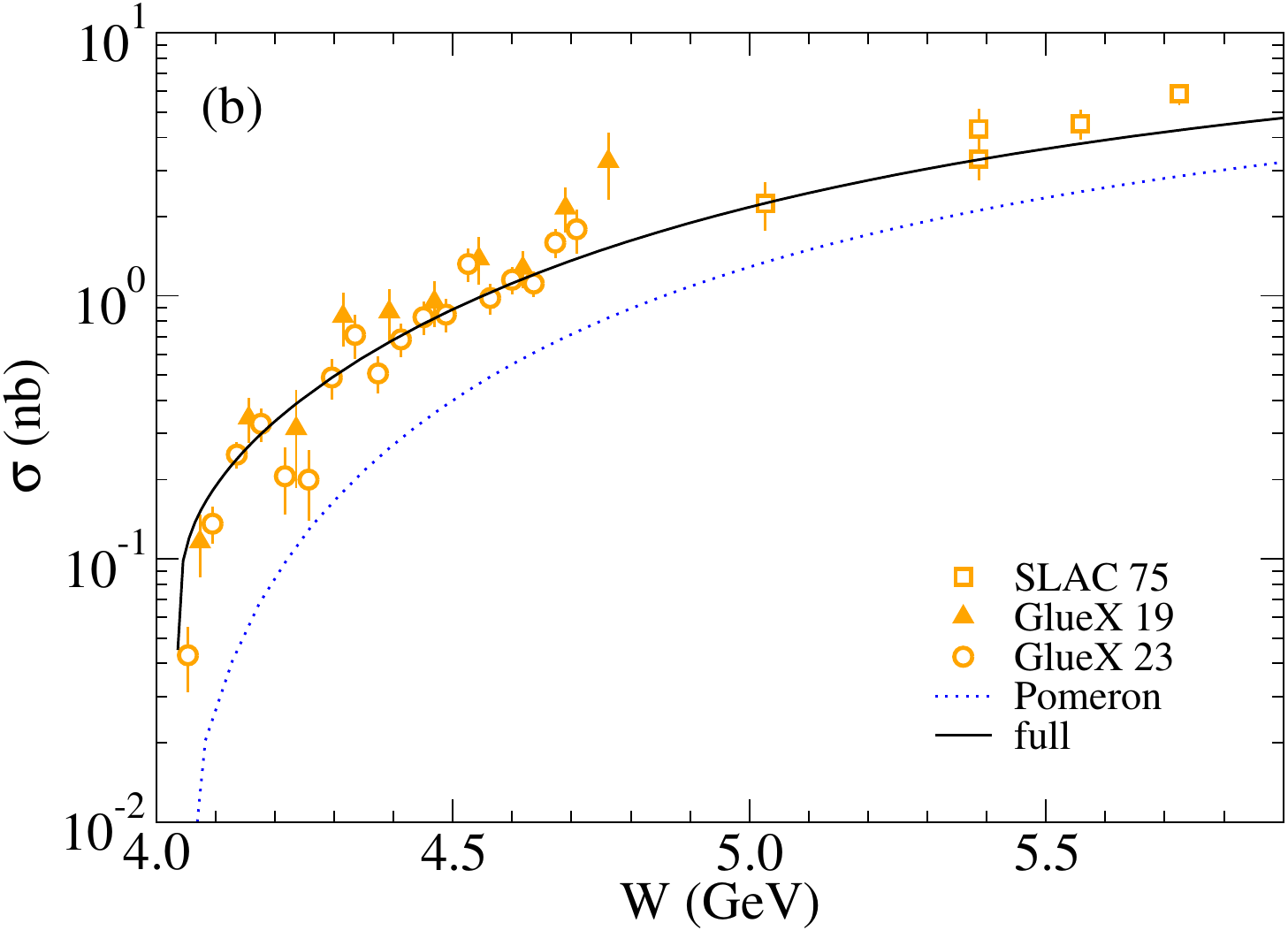}
\caption{(a) Total cross section of $\gamma p \to J/\psi p$ is plotted as a function
of the photon c.m. energy from threshold up to $W = 300$ GeV.
(b) Total cross section at low energies ($4.0 \leqslant W \leqslant 5.9$ GeV).
The SLAC~\cite{Camerini:1975cy}, FermiLab~\cite{Binkley:1981kv,E687:1993hlm},
ZEUS~\cite{ZEUS:2002wfj}, H1~\cite{H1:1996kyo,H1:2000kis}, and
GlueX~\cite{GlueX:2019mkq,GlueX:2023pev} data are compared with the Pomeron-exchange
(dotted curves) and full (solid curves) contributions~\cite{Sakinah:2024cza}.}
\label{fig:totcrst-2y-fit1}
\end{figure}

\section{Formulation of exclusive $J/\psi$ photo-production on nuclei}
\label{Section:III}

The differential cross section of the exclusive photo-production of $J/\psi$ ($V$)
on the nuclear target $A$, $\gamma({\bf q},\lambda) + A({\bf P},M_A ) \to
V({\bf k},m_V) + A({\bf P}^\prime,M^\prime_A)$, in the Laboratory (Lab) frame
(${\mathbf P}=0$), can be written as
\begin{eqnarray}
&& \frac{d\sigma}{d\Omega_{\rm Lab}} =
\rho_A({\bf k},{\bf q})
\frac{1}{2} \frac{1}{2J_A+1}
\sum_{M_A^\prime, M_A} \sum_{m_V,\lambda}   \nonumber \\
&& \hskip 0.3cm \times
| \braket{ {\bf k}m_V,\Phi^{J_A}_{{\bf P}^\prime,M^\prime_A}|T_{VA,\gamma A}(E)|
{\bf q}\lambda,\Phi^{J_A}_{{\bf P}, M_A} } |^2,
\label{eq:dsdt-lab}
\end{eqnarray}
where $\lambda$ is the photon helicity, $J_A$ is the spin of the nucleus $A$,
$M_A$ and $m_V$ are the $z$-components of the spins of $A$ and $V$, respectively,
and
\begin{eqnarray}
\rho_A({\bf k},{\bf q})
&=& \frac{(2\pi)^4 \left\vert \textbf{k} \right\vert^2 E_{V}(\textbf{k})
E_A(\textbf{q}-\textbf{k})}
{\left\vert E_A(\textbf{q} - \textbf{k}) \vert \textbf{k} \vert + E_{V}(\mathbf{k})
( \vert \mathbf{k} \vert - \vert \textbf{q} \vert \cos\theta_{\rm k}) \right\vert}
\,.\nonumber \\
&& \label{eq:rhoa}
\end{eqnarray}
Here ${\bf q}$ is in the $z$-direction and
$\cos\theta_{\rm k}=\hat{\mathbf{q}} \cdot \hat{\mathbf{k}}$.

We will make  predictions of the cross sections of the $\gamma  A \to J/\psi A$
reaction using the multiple scattering theory~\cite{Watson:1960,Kerman:1959fr,
Landau:1973iv,Lee:1974zr,Feshbach:1992}.
In the distorted-wave impulse approximation~\cite{Lee:1974zr}, the reaction
amplitude can be written as
\begin{eqnarray}
T_{VA,\gamma A}(E) = T^{\rm IMP}_{VA,\gamma A}(E) + T^{\rm FSI}_{VA,\gamma A}(E).
\label{eq:ta-tot}
\end{eqnarray}
The impulse (IMP) amplitude is given by
\begin{eqnarray}
T^{\rm IMP}_{VA,\gamma A}(E) = \braket{ \Phi_A|\sum_{i=1,A} t_{VN,\gamma N}(i)|\Phi_A },
\label{eq:imp-0}
\end{eqnarray}
where $\Phi_A$ is the wave function of a nucleus with $A$ nucleons and
$t_{VN,\gamma N}(i)$ is the amplitude of $\gamma N \to J/\psi N$ on the $i$-th nucleon. 
The FSI amplitude takes the form
\begin{eqnarray}
\hskip -0.7cm
T^{\rm FSI}_{VA,\gamma A}(E) = T^{(1)}_{VA,V A}(E)
\frac{1}{E-H_0+i\epsilon} T^{\rm IMP}_{VA,\gamma A}(E),
\label{eq:tfsi}
\end{eqnarray}
where $H_0$ is the free Hamiltonian and $T^{(1)}_{VA,VA}(E)$ is the
$J/\psi A \to J/\psi A$ scattering amplitude calculated from the first-order optical
potential $U^{(1)}_{VA,VA}$ using the following Lippmann-Schwinger equation
\begin{eqnarray}
T^{(1)}_{VA,VA}(E) &=& U^{(1)}_{VA,VA}
\nonumber \\
&+& U^{(1)}_{VA,VA} \frac{1}{E-H_0+i\epsilon}T^{(1)}_{VA,VA}(E).
\label{eq:tvnvn}
\end{eqnarray}
Here the first order optical potential is given by~\cite{Kerman:1959fr,Landau:1973iv}
\begin{eqnarray}
U^{(1)}_{VA,VA} = \braket{ \Phi_A|\sum_{i=1,A}t_{VN,VN}(i)|\Phi_A },
\label{eq:u1}
\end{eqnarray}
where $t_{VN,VN}(i)$ is the amplitude of $J/\psi N \to J/\psi N$ on the $i$-th nucleon.

In Sec.~\ref{Section:IV}, we first consider the simplest nucleus, the deuteron ($d$),
to develop a formulation within which the impulse amplitude of Eq.~(\ref{eq:imp-0})
can be calculated exactly using the deuteron wave function generated from the
realistic nucleon-nucleon ($NN$) potentials~\cite{Wiringa:1994wb,Piarulli:2016vel,
Machleidt:1989tm}.
We then examine the extent to which the approach based on the fixed scatter
approximation (FSA)~\cite{Feshbach:1992,Kerman:1959fr,Landau:1973iv,Lee:1974zr} is
valid for $J/\psi$ photo-production.
In Sec.~\ref{Section:V}, the developed FSA formulation is applied to calculate the
first-order optical potential of Eq.~(\ref{eq:u1}) and the cross sections of the
exclusive $J/\psi$ photo-production on ${^4\rm He}$, ${^{16}\rm O}$, and
${^{40}\rm Ca}$.

\section{Production on the deuteron ($d$) target}
\label{Section:IV}

\subsection{Impulse term}

\begin{figure}[ht] 
\begin{center}
\includegraphics[width=0.70\columnwidth,angle=0]{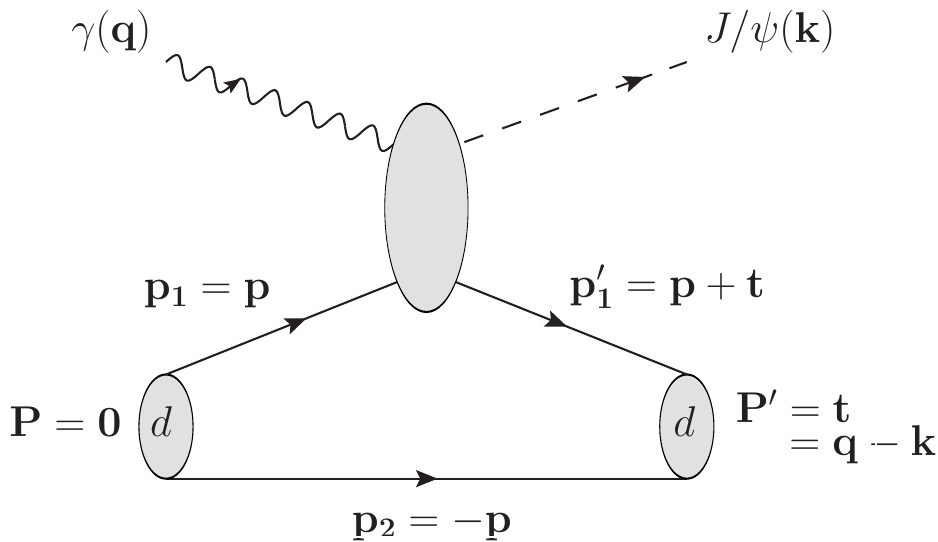}
\caption{Diagrammatic representation of $\gamma d \to J/\psi d$.}
\label{fig:gd-mec}
\end{center}
\end{figure}
We first consider the impulse amplitude in Eq.~(\ref{eq:ta-tot}) for
$\gamma ({\bf q}) + d ({\bf P}) \to J/\psi({\bf k}) + d({\bf P}')$.
The momenta of the involved particles are shown in Fig.~\ref{fig:gd-mec}.
Including the spin quantum numbers, Eq.~(\ref{eq:imp-0}) can be expressed as
\begin{eqnarray}
&& \braket{ {\bf k}m_V,\Phi^{J_d}_{{\bf P}^\prime,M^\prime_d}|T^{\rm IMP}_{Vd,\gamma d}(E)|
{\bf q}\lambda,\Phi^{J_d}_{{\bf P}, M_d} }   \nonumber  \\
&&= \sum_{i=1}^{2} \braket{ \Phi^{J_d}_{{\bf P}^\prime,M^\prime_d}
|\braket{ {\bf k}m_V|t_{VN,\gamma N}(i)|{\bf q}\lambda } |\Phi^{J_d}_{{\bf P}, M_d} },
\,\,\,\,\,\,
\label{eq:imp}
\end{eqnarray}
where $\braket{ {\bf k}m_V|t_{VN,\gamma N}(i)|{\bf q}\lambda }$ is the $J/\psi$
photo-production amplitude on the $i$-th nucleon in the deuteron.
The deuteron wave function is generated from the available $NN$
potentials~\cite{Wiringa:1994wb,Piarulli:2016vel,Machleidt:1989tm} and is defined by
\begin{eqnarray}
&& \Phi^{J_d}_{{\bf P},M_d}({\bf p}_1m_{s_1},{\bf p}_2m_{s_2})   \nonumber\\
&& = \delta ({\bf P}-{\bf p}_1-{\bf p}_2) \Gamma({\bf P}{\bf p},{\bf p}_1{\bf p}_2)
\phi^{J_d}_{M_d}({\bf p},m_{s_1}m_{s_2}),
\,\,\,\,\,\,\,\,\,\,\,
\label{eq:deutf}
\end{eqnarray}
where $m_{s_i}$ is the $z$-component of the spin of the $i$-th nucleon, ${\bf P}$ is
the total momentum of $d$, ${\bf p}_i$ is the momentum of the $i$-th nucleon, and
$\Gamma({\bf P}{\bf p},{\bf p}_1{\bf p}_2)$ is the transformation coefficient from
the momenta $({\bf p}_1,{\bf p}_2)$ in the Lab frame to $({\bf P},{\bf p})$ in
the rest frame of the deuteron. 
The momentum variables in Eq.~(\ref{eq:deutf}) are explicitly defined by
\begin{eqnarray}
{\bf P} &=& {\bf p}_1+{\bf p}_2,
\label{eq:bigp}\\
{\bf p} &=& \frac{1}{2}({\bf p}_1-{\bf p}_2),
\label{eq:smap}\\
\Gamma({\bf P}{\bf p},{\bf p}_1{\bf p}_2)&=&
\braket{ {\bf p}_1{\bf p}_2|{\bf P}{\bf p} }
\nonumber \\
&=& \sqrt{ \frac{\partial({\bf P},{\bf p})}{\partial({\bf p}_1,{\bf  p}_2)} }.
\label{eq:jacob}
\end{eqnarray}
With the nonrelativistic kinematics in Eqs.~(\ref{eq:bigp})-(\ref{eq:smap}), the
transformation coefficient of Eq.~(\ref{eq:jacob}) obviously becomes
\begin{eqnarray}
\Gamma({\bf P}{\bf  p},{\bf p}_1{\bf p}_2)&=&1.
\label{eq:tr-non}
\end{eqnarray}
In Eq.~(\ref{eq:deutf}), $\phi^{J_d}_{M_d}({\bf p},m_{s_1}m_{s_2})$ is the wave function
in the rest frame (${\bf P}=0$) of the deuteron and takes the form
\begin{eqnarray}
&& \phi^{J_d}_{M_d}({\bf p},m_{s_1} m_{s_2}) =
\sum_{L,M_L,M_S} \braket{ J_d M_d|L S M_L M_S }   \nonumber \\
&& \hskip 1.0cm \times
\braket{ SM_S| \textstyle\frac12 \textstyle\frac12 m_{s_1}m_{s_2} }
Y_{LM_L}(\hat{p}) R_L(|{\bf p}|),
\end{eqnarray}
where $J_d=1$, $S=1$, $L$ is the relative orbital angular momentum of the
two nucleons in $d$, and $R_L(|{\bf p}|)$ is the radial part of the deuteron
wave function.

Using the deuteron wave function given in Eq.~(\ref{eq:deutf}) and noting that 
\begin{eqnarray}
\Gamma({\bf 0}{\bf p},{\bf p}_1{\bf p}_2) = 1,
\end{eqnarray}
Eq.~(\ref{eq:imp}) can be written as
\begin{eqnarray}
&& \braket{ {\bf k}m_V,\Phi^{J_d}_{{\bf P}^\prime,M^\prime_d}
|T^{\rm IMP}_{Vd,\gamma d}(E)|{\bf q}\lambda, \Phi^{J_d}_{{\bf P},M_d} } = \nonumber \\
&& \sum_{m_{s_1},m_{s_2},m_{s'_1},m_{s'_2}} A_d \int d{\bf p}\,
\phi^{J_d *}_{M'_d}({\bf p}',m_{s'_1}m_{s'_2})   \nonumber \\
&& \times \Gamma({\bf P}'{\bf p}',{\bf p}'_1{\bf p}'_2)
\phi^{J_d }_{M_d}({\bf p},m_{s_1}m_{s_2})   \nonumber \\
&& \times \braket{ {\bf k}m_V,{\bf p}'_1m_{s'_1}|t_{VN,\gamma N}(\omega)|
{\bf q}\lambda, {\bf p}m_{s_1} },
\label{eq:gd-amp-0}
\end{eqnarray}
where $A_d=2$ is the number of the nucleons in the deuteron.
As illustrated in Fig.~\ref{fig:gd-mec}, the momentum variables in
Eq.~(\ref{eq:gd-amp-0}) are defined in the Lab frame by 
\begin{eqnarray}
&& {\bf p}_1={\bf p};\, {\bf p}_2=-{\bf p},   \\
&&{\bf p}'_1={\bf p}+{\bf t};\, {\bf p}'_2=-{\bf p},
\end{eqnarray}
where
${\bf t}={\bf q}-{\bf k}$
and
\begin{eqnarray}
{\bf p}'=\frac{1}{2}({\bf p}'_1-{\bf p}'_2)={\bf p}+\frac{1}{2}{\bf t}\,.
\label{eq:smapp}
\end{eqnarray}
By setting $\Gamma({\bf P}'{\bf p}',{\bf p}'_1{\bf p}'_2)=1$, as given in
Eq.~(\ref{eq:tr-non}), Eqs.~(\ref{eq:gd-amp-0})-(\ref{eq:smapp})
are for the nonrelativistic calculations of the impulse amplitude.

The main feature of the $J/\psi$ photo-production at energies near the threshold is
that the momentum transfer $-t$ is very large.
For example, the contributions to the cross sections at photon energy $E_\gamma = 6$
GeV are from the $-t \geqslant$ 2 GeV$^2$ region where the relativistic effects are
expected to be important.
We therefore follow
Refs.~\cite{Kamada:2002qt,Witala:2004pv,Witala:2008va,Grassi:2022flz,Wu:2013xma} 
to modify the above nonrelativistic formula to account for relativistic effects using
the instant form of Relatvistic Quantum Mechanics~\cite{Dirac:1949cp,Keister:1991sb}.

The main step is to replace the relative momentum ${\bf p}$ of Eq.~(\ref{eq:smap}) by
the following expression
\begin{eqnarray}
{\bf p}= L(\vec{\beta}) {\bf p}_1 = -L(\vec{\beta}) {\bf p}_2,
\label{eq:smap-a}
\end{eqnarray}
where $L(\vec{\beta})$ is the Lorentz Boost transformation with the velocity defined
by
\begin{eqnarray}
\vec{\beta}=\frac{{\bf P}}{E_N({\bf p}_1)+E_N({\bf p}_2)}.
\end{eqnarray}
It can be shown~\cite{Keister:1991sb,Wu:2013xma} that for a particle with momentum
${\bf p}_i$ and mass $m$
\begin{eqnarray}
L(\vec{\beta}) {\bf  p}_i= {\bf p}_i +
\gamma \left[ \frac{\gamma}{1+\gamma}(\vec{\beta}\cdot {\bf p}_i)-E_m({\bf p}_i)
\right],
\label{eq:smap-b}
\end{eqnarray}
where $\gamma =1/ \sqrt{1-\vec{\beta}^{\,2}}$ and
$E_m({\bf p}_i)=\sqrt{{\bf p}_i^2+m^2}$.
Using ${\bf P}$ of Eq.~(\ref{eq:bigp}) and ${\bf p}$ of
Eqs.~(\ref{eq:smap-a})-(\ref{eq:smap-b}), one can show that~\cite{Keister:1991sb}
the transformation coefficient of Eq.~(\ref{eq:jacob}) becomes
\begin{eqnarray}
\hskip -0.6cm
\Gamma({\bf P}{\bf p},{\bf p}_1{\bf p}_2)
= \left[ \frac{E_1({\bf p}_1)+E_2({\bf p}_2)}{E_1({\bf p})+E_2({\bf p})}
\frac{E_1({\bf p})E_2({\bf p})}{E_1({\bf p_1})E_2({\bf p}_2)} \right]^{\frac12}.
\label{eq:jacob-rel}
\end{eqnarray}
Relativistic effects can be included using the same equations of
Eqs.~(\ref{eq:gd-amp-0})-(\ref{eq:smapp}) but using Eq.~(\ref{eq:jacob-rel}) to
evaluate $\Gamma({\bf P}'{\bf p}',{\bf p}'_1{\bf p}'_2)$ and replacing ${\bf p}'$
of Eq.~(\ref{eq:smapp}) by 
\begin{eqnarray}
{\bf p}'=L(\vec{\beta}^{\,\prime}) {\bf  p}'_1 =
-L(\vec{\beta}^{\,\prime}){\bf  p}'_2,
\label{eq:rel-p-a}
\end{eqnarray}
where
\begin{eqnarray}
\vec{\beta}^{\,\prime}=\frac{{\bf t}}{E_N({\bf p}'_1)+E_N({\bf p}'_2)}\,.
\label{eq:rel-p-b}
\end{eqnarray}

In addition to the above relativistic extension, the spin rotaions under Lorentz
Boost transformation are also needed to have a complete relativistic approach.
However, it was found in Refs.~\cite{Kamada:2002qt,Witala:2004pv,Wu:2013xma} that the
spin rotations have negligible effects on the spin-averaged observables which are only
considered in this work. 
Thus we will not account for this complication for simplicity.

\begin{figure}[h] 
\begin{center}
\includegraphics[width=0.80\columnwidth,angle=0]{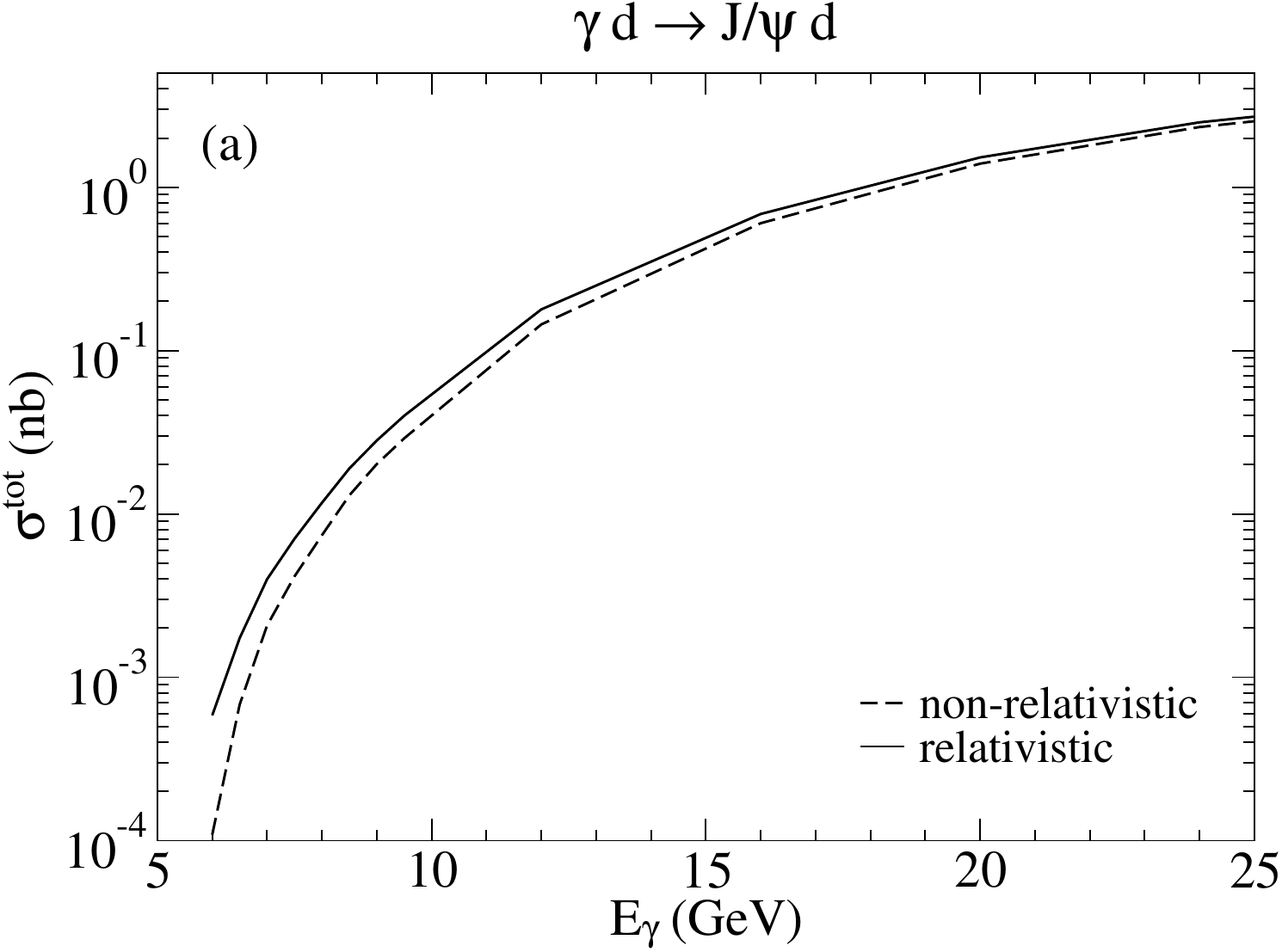} \\
\vspace{0.5em}
\includegraphics[width=0.80\columnwidth,angle=0]{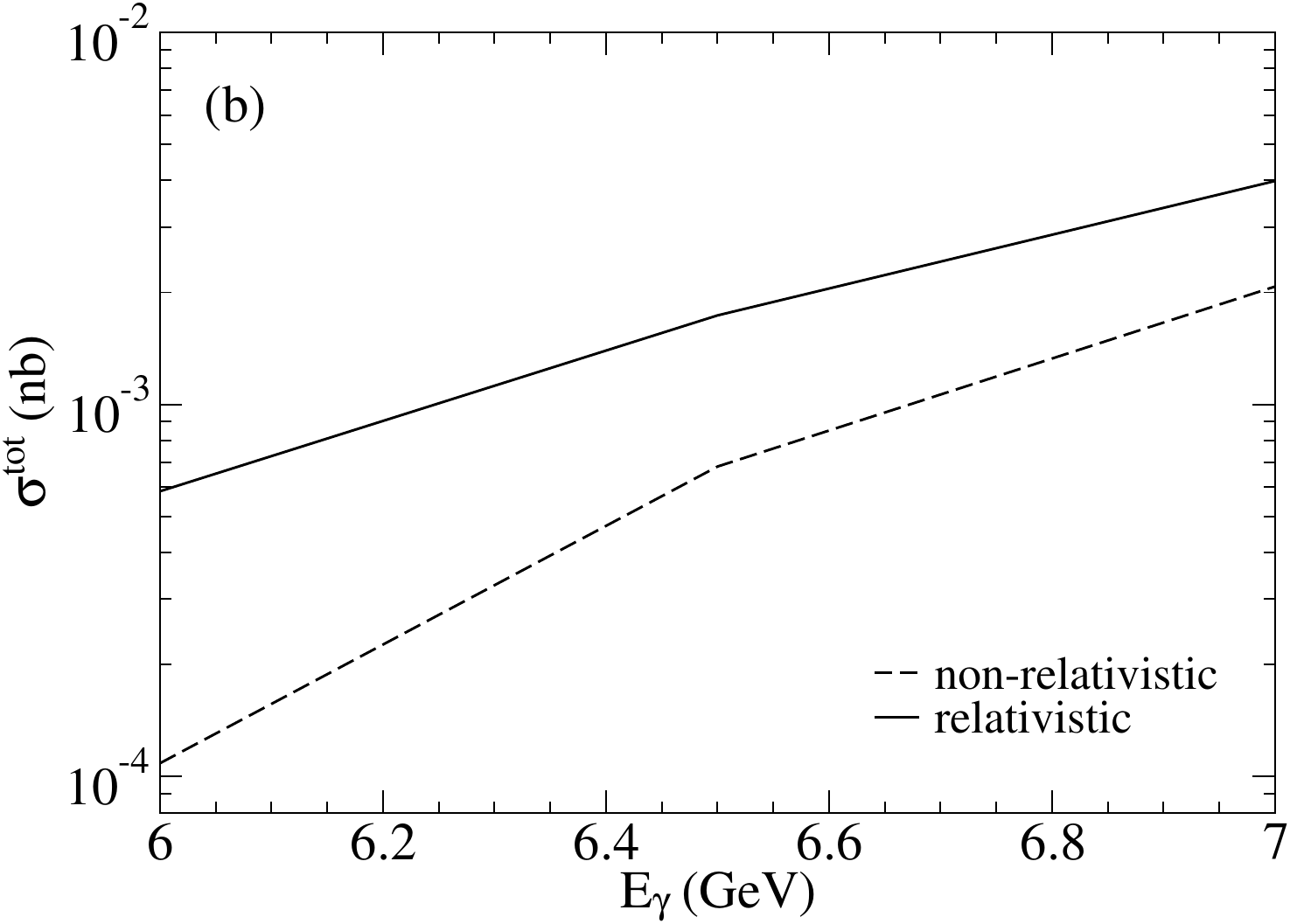}
\caption{(a) Total cross section of $\gamma d \to J/\psi d$ is plotted as a function
of the photon Lab energy from threshold up to $E_\gamma = 25$ GeV.
(b) Total cross section at low energies ($6 \leqslant E_\gamma \leqslant 7$ GeV).
The results using the nonrelativistic (dashed curves) and relativistic (solid curves)
formulation of the deuteron wave function are compared.}
\label{fig:totcrst-rel-nonr}
\end{center}
\end{figure}

The resulting total cross section of $\gamma d \to J/\psi d$ is depicted in
Fig.~\ref{fig:totcrst-rel-nonr} as a function of $E_\gamma$ (a) from threshold up to
$E_\gamma = 25$ GeV and (b) at low energies ($6 \leqslant E_\gamma \leqslant 7$ GeV).
The results using the nonrelativistic and relativistic formulation of the deuteron
wave function are compared to each other.
We see that their differences are very large near the threshold region and are
much less at higher energies.

\begin{figure}[t] 
\begin{center}
\includegraphics[width=0.80\columnwidth,angle=0]{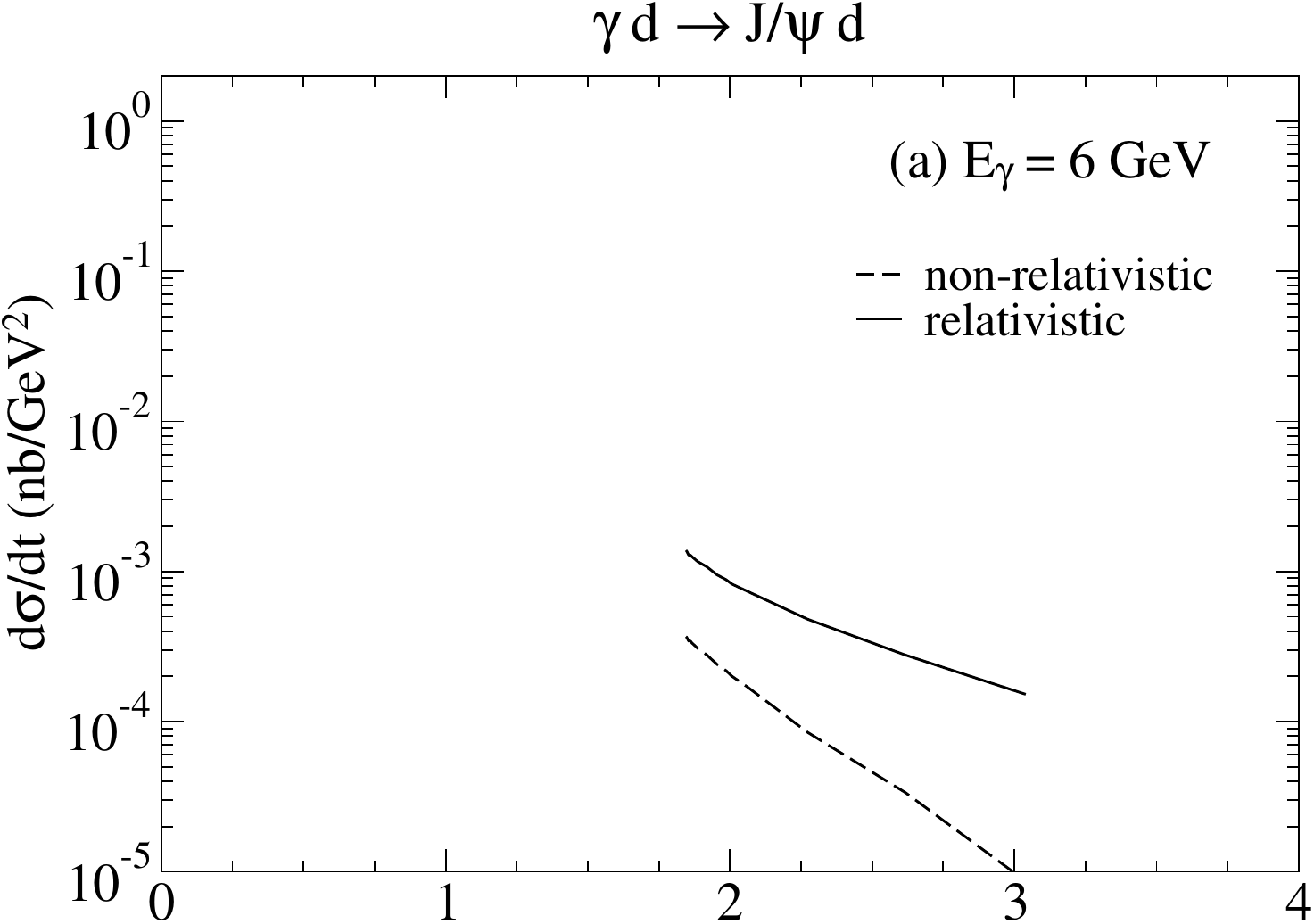} \\
\vspace{0.3em}
\includegraphics[width=0.80\columnwidth,angle=0]{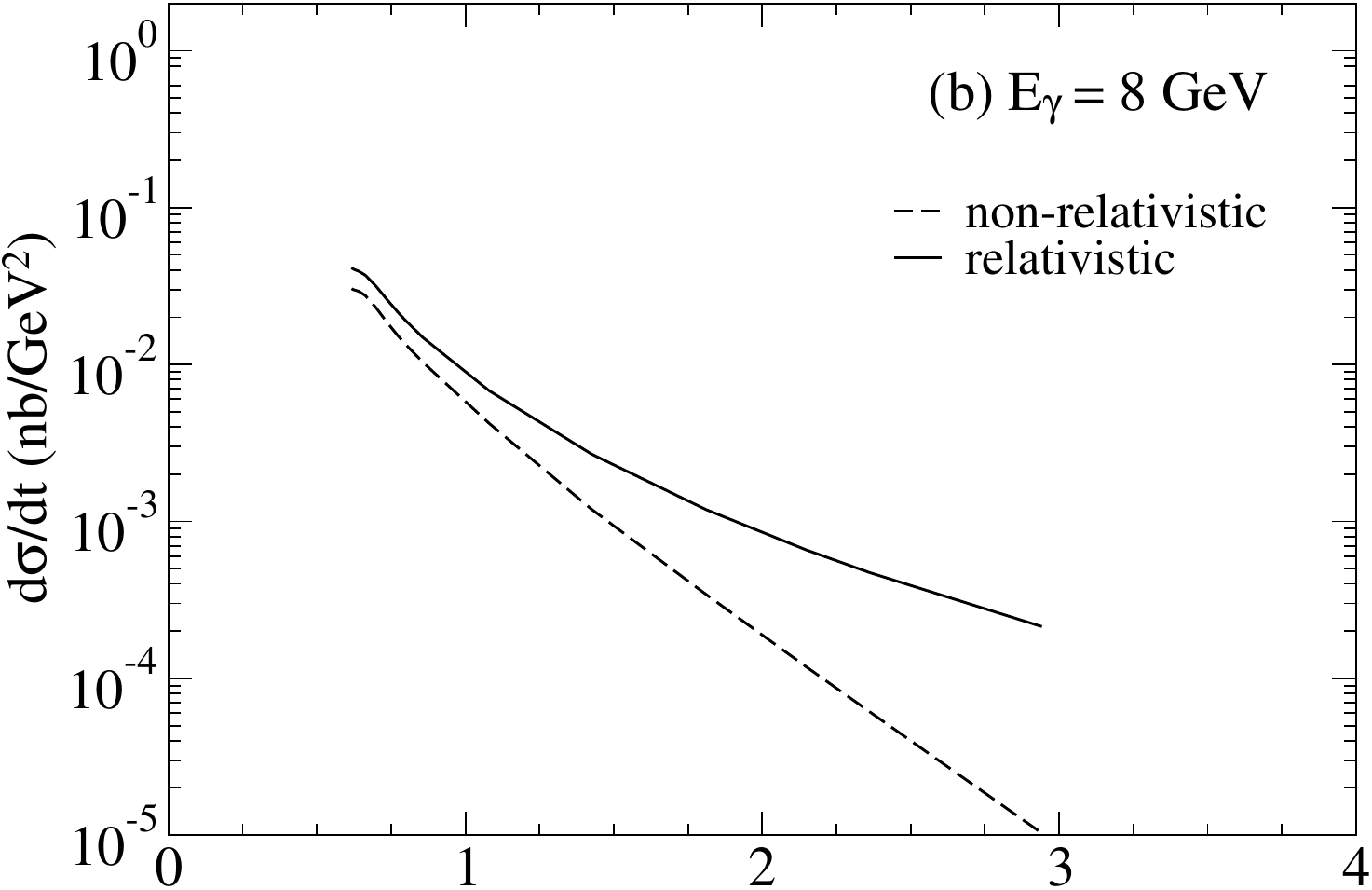} \\
\vspace{0.3em}
\includegraphics[width=0.80\columnwidth,angle=0]{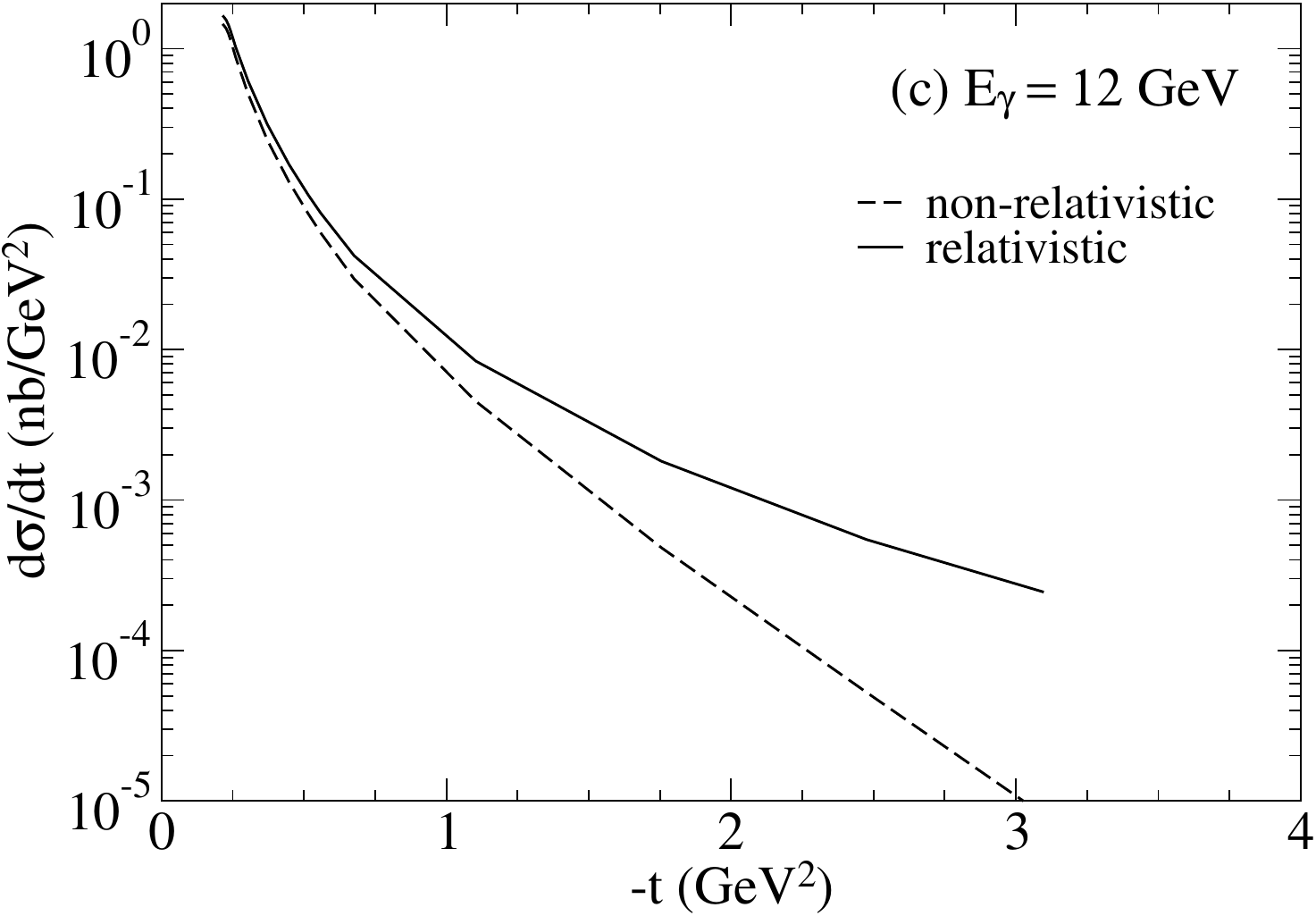}
\caption {Differential cross sections of $\gamma d \to J/\psi d$ at
$E_\gamma =$ (6, 8, 10) GeV.
The results using the nonrelativistic (dashed curves) and relativistic (solid curves)
formulation of the deuteron wave function are compared.}
\label{fig:dsdt-rel-nonr}
\end{center}
\end{figure}

This can be understood from Fig.~\ref{fig:dsdt-rel-nonr} where the differential
cross sections are presented as functions of $-t$.
At $E_\gamma = 6$ GeV (a), which is about 0.4 GeV above the threshold, the results
using the relativistic deuteron wavefunction are about an order of magnitude larger
than those using the nonrelativistic one in the large momentum-transfer region of
$-t \geqslant 1.8$ GeV$^2$.
As the photon energy increases, the total cross sections are mainly from the low
momentum-transfer region where the relativistic effects are not large as seen in
Fig.~\ref{fig:dsdt-rel-nonr}(b)-(c).

\subsection{Fixed scatter approximation (FSA)}

For understanding the results of differential cross sections shown in
Fig.~\ref{fig:dsdt-rel-nonr}, it is useful to use an approximation to relate
Eq.~(\ref{eq:gd-amp-0}) to the nuclear form factors which characterize the structure
of the deuteron.
This can be done using the FSA~\cite{Feshbach:1992,Kerman:1959fr,Landau:1973iv,
Lee:1974zr} to set the initial nucleon momentum to be ${\bf p}=0$ in the amplitude
$\braket{ {\bf k}m_V,{\bf p}'_1m_{s'_1}|t_{VN,\gamma N}(\omega)|{\bf q}\lambda,
{\bf p}m_{s_1} }$ and also to set the transformation
coefficient $\Gamma({\bf P}'{\bf p}',{\bf p}'_1{\bf p}'_2)=1$.
Eq.~(\ref{eq:gd-amp-0}) can then be approximately written as the following factorized
form   
\begin{eqnarray}
&& \hskip -0.6cm
\braket{ {\bf k}m_V,\Phi^{J_d}_{{\bf P}^\prime,M^\prime_d}|T^{\rm IMP}_{Vd,\gamma d}(E)|
{\bf q}\lambda,\Phi^{J_d}_{{\bf P}, M_d} } \sim \nonumber \\
&& \hskip -0.6cm
A_d \braket{ {\bf k}m_V,{\bf t}\bar{m}_{s'_1}|{\bar t}_{VN,\gamma N}(\omega_0)
|{\bf q}\lambda,{\bf 0}\bar{m}_{s_1} }
F_{M'_d,M_d}({ t}),
\label{eq:amp-fact-d}
\end{eqnarray}
where $\bar{m}_{s_1}$ and $\bar{m}_{s'_1}$ are arbitrarily chosen spin components,
$t=(|{\bf q}|-E_V({\bf k}))^2-{\bf t}^2$, and
\begin{eqnarray}
&& \hskip -0.6cm
F_{M'_d,M_d}({t}) =
\sum_{m_{s_1},m_{s_2},m_{s'_1},m_{s'_2}}
\nonumber \\
&& \hskip -0.6cm
\times \int d{\bf p}\,
\phi^{J_d *}_{M'_d}({\bf p}+\textstyle\frac{{\bf t}}{2},m_{s'_1}m_{s'_2})
\,\phi^{J_d}_{M_d}({\bf p},m_{s_1}m_{s_2}).
\label{eq:d-ff}
\end{eqnarray}

\begin{figure}[h] 
\begin{center}
\includegraphics[width=0.80\columnwidth,angle=0]{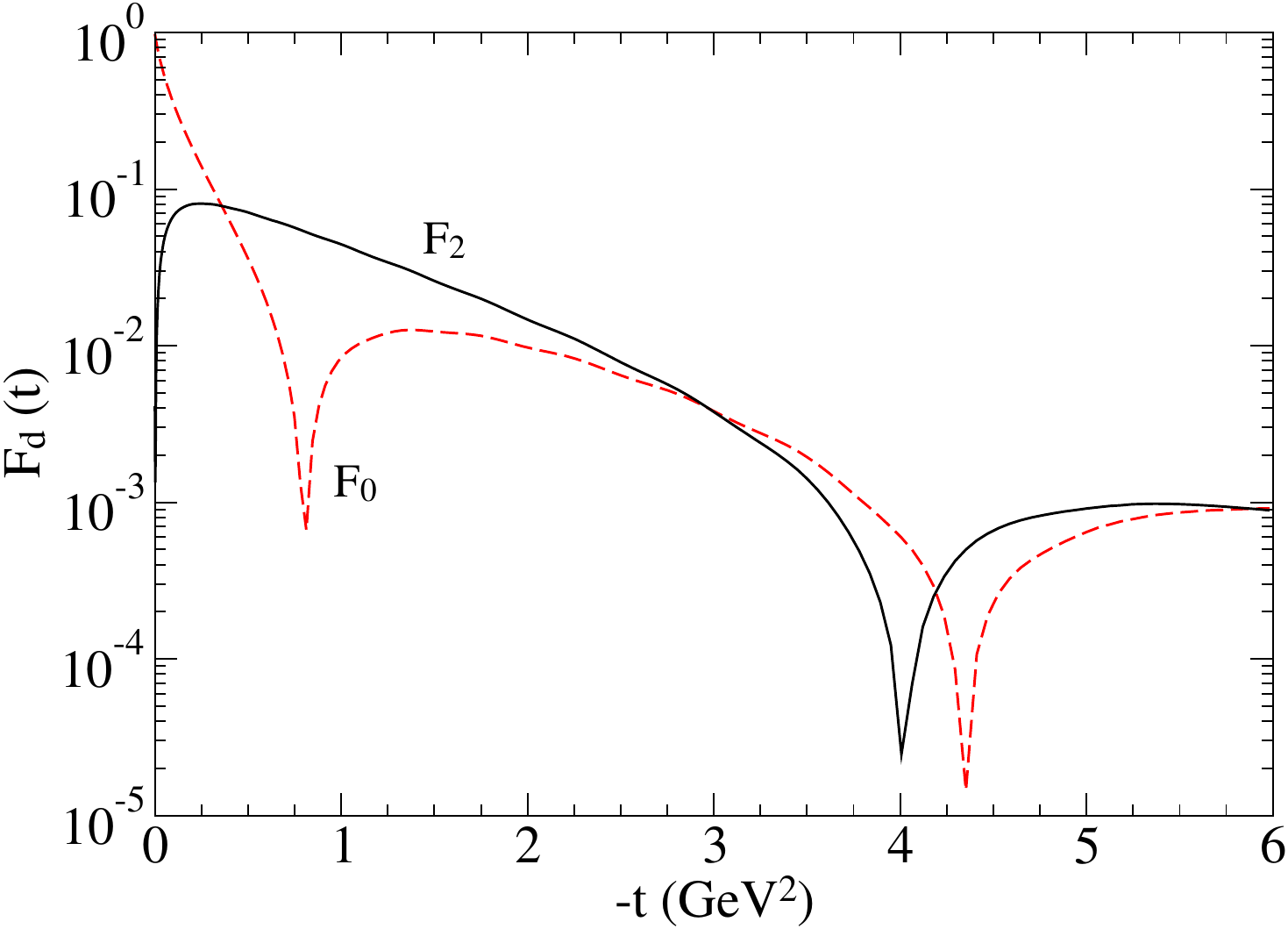}
\caption{Deuteron form factors calculated using the deuteron wave function
from the Argonne-V18 $NN$ potential~\cite{Wiringa:1994wb}.}
\label{fig:deut-f0-f2}
\end{center}
\end{figure}

By choosing ${\bf t}$ in the $z$-direction, one can show that 
\begin{eqnarray}
&& F_{0,0}({t}) = \sqrt{4\pi}[F_0({t})-\sqrt{2} F_2({ t})] \label{eq:fmmp-1},
\nonumber \\
&& F_{1,1}({t}) = F_{-1,-1}({t})=
\sqrt{4\pi}
   [F_0({t})+\textstyle\frac{1}{\sqrt{2}} F_2({ t})] \label{eq:fmmp-2},
\nonumber \\
&&F_{M'_d,M_d}({t}) = 0\,\,\, {\rm if} \,\,\,M'_d\neq M_d,
\label{eq:fmmp-3}
\end{eqnarray}
where $F_L(t)$ with $L=(0,2)$ are  the deuteron form factors.
Eqs.~(\ref{eq:amp-fact-d})-(\ref{eq:fmmp-3}) imply that the reaction cross sections
are closely related to the deuteron form factors.
In Fig.~\ref{fig:deut-f0-f2}, we show the deuteron form factors generated from the
Argonne-V18 potential~\cite{Wiringa:1994wb}.
We see that the quadrupole form factor $F_2(t)$ peaks in the region where $F_0(t)$
has a minimum.
It is clear that the reaction cross sections at large $-t$ depend strongly on
$F_2(t)$ which is due to the $d$-state of the wave function.

\begin{figure}[ht] 
\begin{center}
\includegraphics[width=0.80\columnwidth,angle=0]{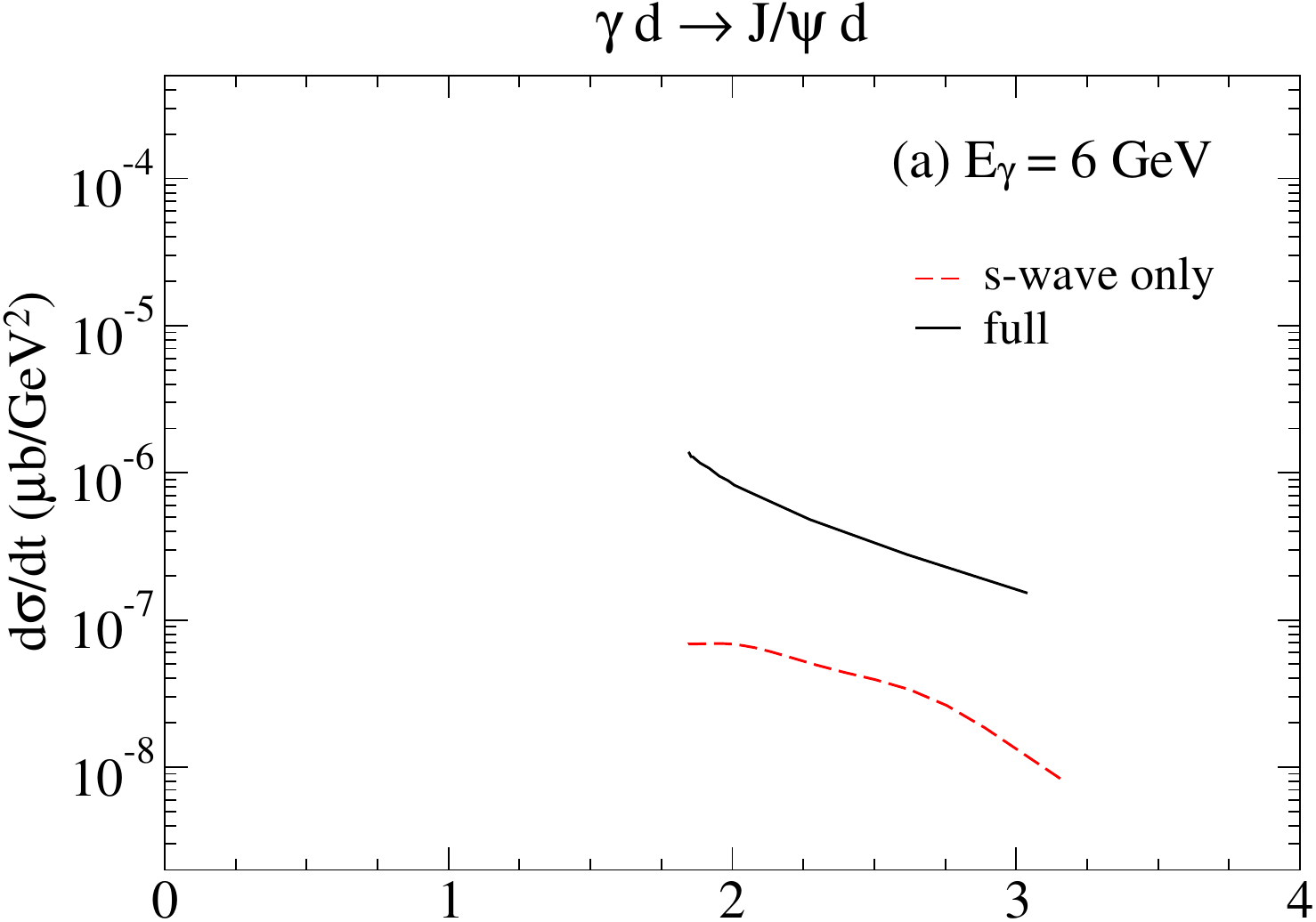} \\
\vspace{0.3em}
\includegraphics[width=0.80\columnwidth,angle=0]{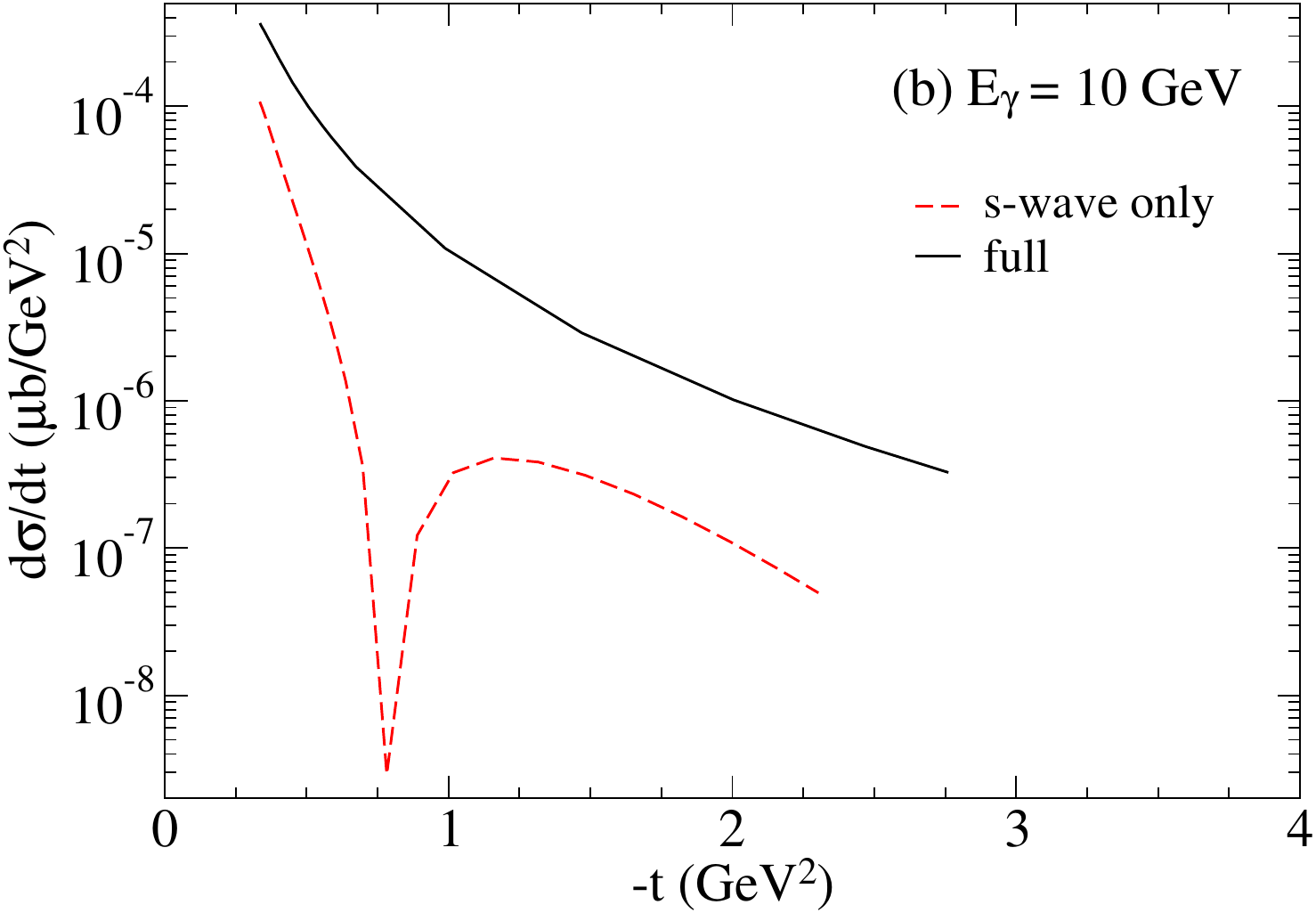}
\caption{Differential cross sections of $\gamma d \to J/\psi d$ at
$E_\gamma =$ (6, 10) GeV.
The results using only the $s$-wave part (dashed curves) and using all parts
(solid curves) of the deuteron wave function are compared.
(Note that the unit for $d\sigma/dt$ in this figure is in $\rm{\mu b/ GeV^2}$.)}
\label{fig:gd-pomcqm-dsdt-s}
\end{center}
\end{figure}

The dependence of the cross sections calculated using the relativistic form
of Eq.~(\ref{eq:gd-amp-0}) is expected to be similar to that discussed above using the
FSA of Eqs.~(\ref{eq:amp-fact-d})-(\ref{eq:fmmp-3}).
To see how the $d$-state can be probed by the $J/\psi$ exclusive production process, 
we compare in Fig.~\ref{fig:gd-pomcqm-dsdt-s} the full results, which correspond to
the solid curves in Fig.~\ref{fig:dsdt-rel-nonr}, with
results obtained by keeping only the $s$-wave part of the deuteron wave function in
the calculations using the relativistic form of Eq.~(\ref{eq:gd-amp-0}).
At $E_\gamma = 6$ GeV (a) which is close to the threshold, both results of
$d\sigma/dt$ have no minimum because they only cover the region of large $-t$ where
both $F_0(t)$ and $F_2(t)$ have no minimum.
In the same $-t$ region, $F_2(t)$ is larger than $F_0(t)$ and hence the results from
keeping only the $s$-wave part of the deuteron wavefunction are much smaller.
At higher energies, the small $-t$ region is covered and the cross sections from
only the $s$-wave part of the deuteron wavefunction have a minimum similar to that of
$F_0(t)$ as displayed in Fig.~\ref{fig:gd-pomcqm-dsdt-s}(b).
Thus the production near the threshold is very effective in testing the $d$-state of
the deuteron wave function.

\begin{figure}[t] 
\begin{center}
\includegraphics[width=0.80\columnwidth,angle=0]{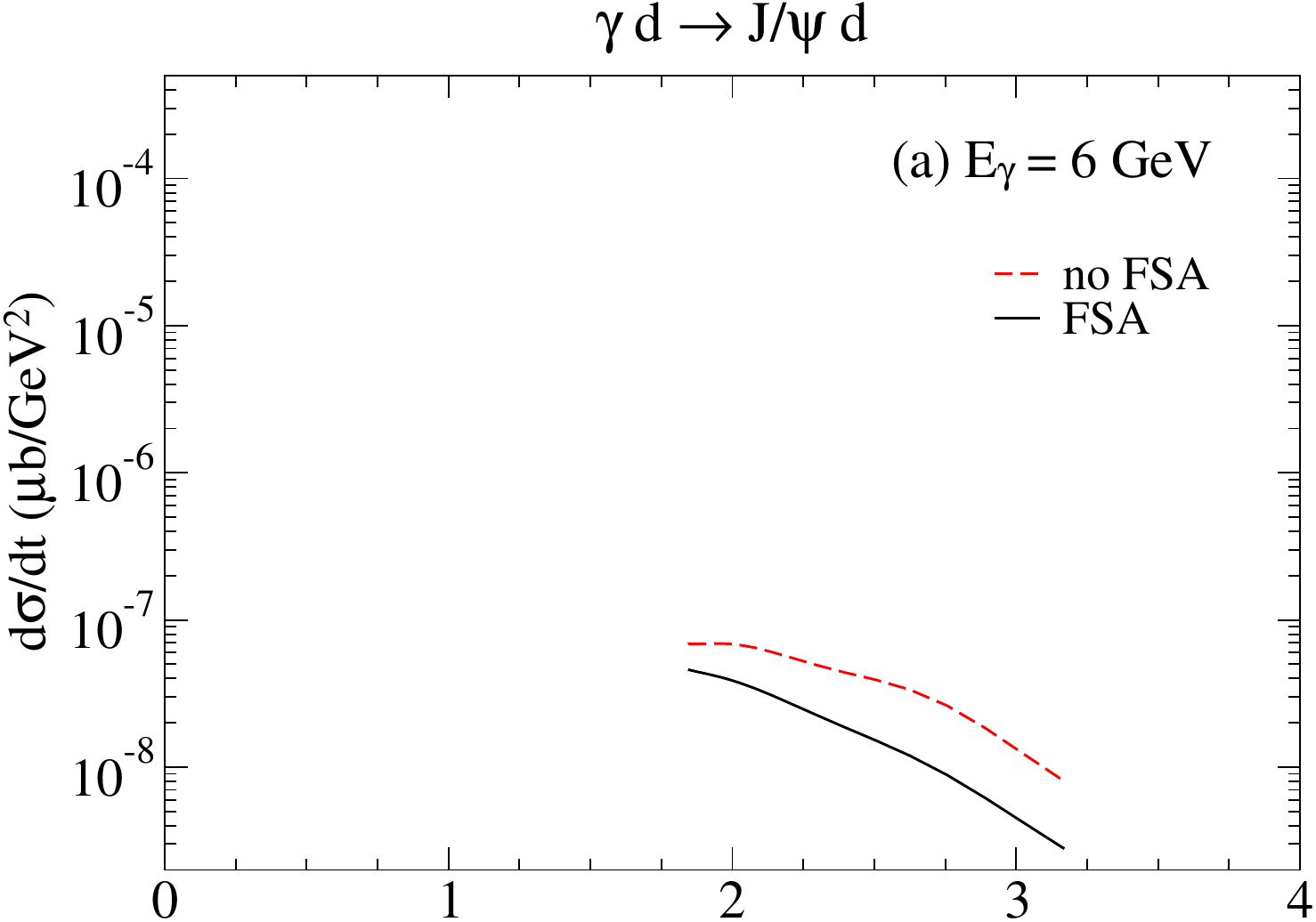} \\
\vspace{0.3em}
\includegraphics[width=0.80\columnwidth,angle=0]{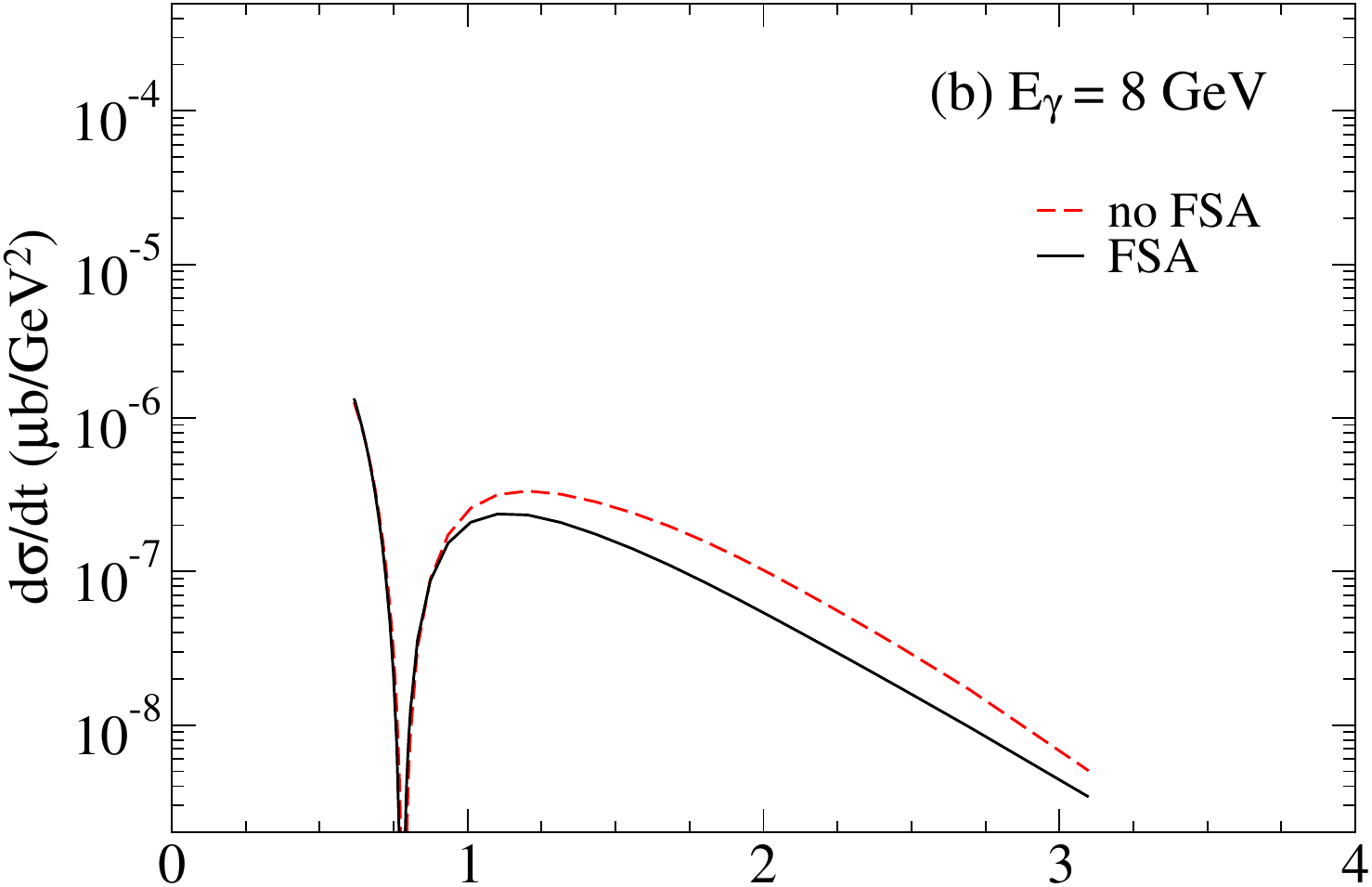} \\
\vspace{0.3em}
\includegraphics[width=0.80\columnwidth,angle=0]{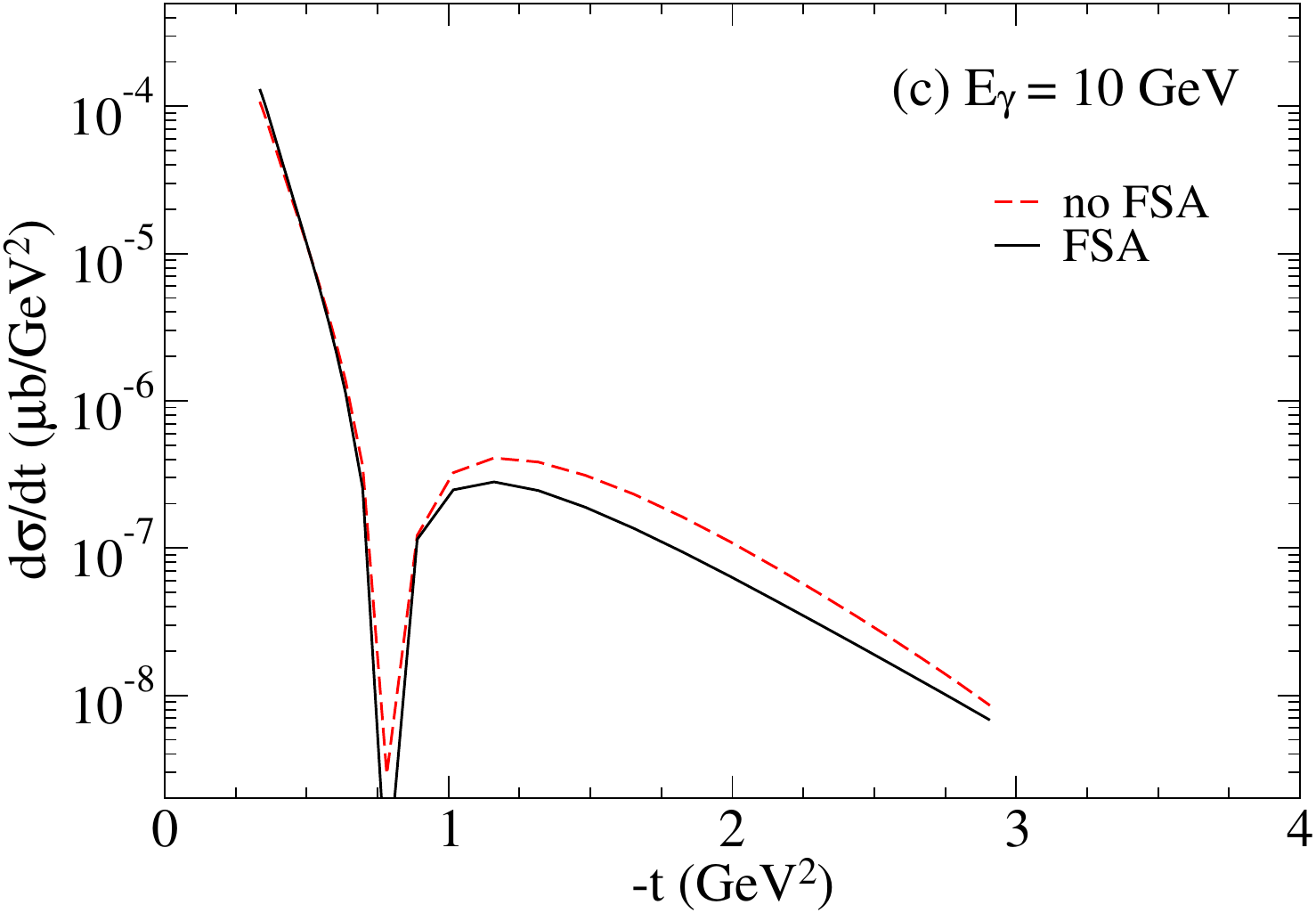}
\caption{Differential cross sections of $\gamma d \to J/\psi d$ at $E_\gamma =$
(6, 8, 10) GeV.
The results using the FSA (solid curves) are compared with the exact results (dashed
curves).
(Note that the unit for $d\sigma/dt$ in this figure is in $\rm{\mu b/ GeV^2}$.)}
\label{fig:check-fsa}
\end{center}
\end{figure}

The FSA is useful in practice and will be used later in calculating the FSI term and
the cross sections for the production on heavier nuclei.
Therefore we need to examine the extent to which it is valid.
The results can be understood clearly in calculations keeping only the $s$-wave part
of the deuteron wavefunction, such that the cross sections only depend on one form
factor $F_0(t)$.
Using the relativistic form, the FSA results can be obtained by simply setting the
initial nucleon momentum ${\bf p}=0$ in Eq.~(\ref{eq:gd-amp-0}).
The resulting differential cross sections (solid curves) are depicted in
Fig.~\ref{fig:check-fsa} where the dashed curves correspond to the dashed curves in
Fig.~\ref{fig:gd-pomcqm-dsdt-s}.
The FSA is found to be a good approximation only in the small $-t$ region.
Thus the FSA is not valid at energies near the threshold where the cross sections are
only from the large $-t$ region, as shown in Fig.~\ref{fig:check-fsa}(a).
On the other hand, the FSA is reasonable at higher energies $E_\gamma=$ 8 (b) and 10
(c) GeV.

The FSA is employed in Ref.~\cite{Kim:2021adl} for $\phi$ photo-production on the spin
$J = 0$ nucleus $^4\rm{He}$ and will be used in our later calculations for heavier
nuclei. 
It is therefore useful to see how it can be derived from Eq.~(\ref{eq:amp-fact-d}).
To proceed, we need to first specify the undefined spin components $\bar{m}_s$ and
$\bar{m}'_s$ in Eq.~(\ref{eq:amp-fact-d}).
It is reasonable to follow the expression for the production cross sections on the
nucleon to define the averaged amplitude in Eq.~(\ref{eq:amp-fact-d}) as
\begin{eqnarray}
&& \hskip -0.8cm
| \braket{ {\bf k} m_V, {\bf t} \bar{m}_{s'_1} | \bar{t}_{Vd,\gamma d} (\omega_0)
| {\bf q} \lambda, {\bf 0} \bar{m}_{s_1} } |^2
\nonumber \\
&& \hskip -0.8cm
= \frac{1}{2}
\sum_{m_{s'_1}, m_{s_1}}
| \braket{ {\bf k} m_V, {\bf t} m_{s'_1} | t_{VN,\gamma N} (\omega)
| {\bf q} \lambda, {\bf 0} m_{s_1} } |^2.
\label{eq:amp-fact}
\end{eqnarray}
This allows us to cast Eq.~(\ref{eq:dsdt-lab}) into the following FSA form
\begin{eqnarray}
\left( \frac{d\sigma}{d\Omega_{\rm Lab}} \right)_{\rm FSA} &=&
\rho_d({\bf k},{\bf q})
\frac{1}{4} \sum_{m_{s'_1},m_{s_1},m_V,\lambda}
\nonumber \\
&& \hskip -1.7cm
\times |\braket{ {\bf k}m_V,{\bf t}{m}_{s'_1}|T^{\rm FSA}_{Vd,\gamma d}(\omega)
|{\bf q}\lambda,{\bf 0}{m}_{s_1} } |^2,
\label{eq:cross-fsa}
\end{eqnarray}
where
\begin{eqnarray}
&&\braket{ {\bf k}m_V,{\bf t}{m}_{s'_1}|T^{\rm FSA}_{Vd,\gamma d}(\omega)
|{\bf q}\lambda,{\bf 0}{m}_{s_1} }\nonumber \\
&& = A_d \braket{ {\bf k}m_V,{\bf t}{m}_{s'_1}|{t}_{VN,\gamma N}(\omega)
|{\bf q}\lambda,{\bf 0}{m}_{s_1} }
F_{ave}(t).
\label{eq:t-fsa}
\end{eqnarray}
Here we used the properties of Eq.~(\ref{eq:fmmp-1}) to define
\begin{eqnarray}
F_{ave}(t) &=&\frac{1}{\sqrt{4\pi}} \left[\frac{1}{2J_d+1}
\sum_{M'_d,M_d} |F_{M'_d,M_d}|^2 \right]^{\frac12} \nonumber \\
&=& [F_0({t})|^2+F_2({t})^2]^{\frac12}.
\label{eq:fave}
\end{eqnarray}
The FSA defined by Eqs.~(\ref{eq:cross-fsa})-(\ref{eq:t-fsa}) is the same as the
formulation employed in Ref.~\cite{Kim:2021adl} for $\phi$ photo-production on the
spin $J = 0$
nucleus $^4\rm{He}$.



\subsection{Final state interactions}

We now turn to investigating the effects of the FSI term $T^{\rm FSI}_{V d, \gamma d} (E)$
in Eq.~(\ref{eq:ta-tot}). 
As defined in Eq.~(\ref{eq:tfsi}), we need to evaluate the following matrix element
\begin{eqnarray}
&& \braket{ {\bf k}m_V,\Phi^{J_d}_{{\bf P}^\prime,M^\prime_d}|T^{\rm FSI}_{Vd,\gamma d}(E)|
{\bf q}\lambda,\Phi^{J_d}_{{\bf P}, M_d} } =  \nonumber \\
&&\sum_{\bar{m}_V,\bar{M}_d}
\int d{\bf k}^\prime
\braket{ {\bf k}m_V,\Phi^{J_d}_{{\bf P}^\prime,M^\prime_d}|T^{(1)}_{Vd,Vd}(E)|
{\bf k}^\prime \bar{m}_V,\Phi^{J_d}_{{\bf \bar{P}},\bar{M}_d} }   \nonumber \\
&& \times \frac{1}{E-E_V(k')-E_d(\bar{P})+i\epsilon}   \nonumber \\
&& \times \braket{ {\bf k}'\bar{m}_V,\Phi^{J_d}_{{\bf \bar{P}},\bar{M}_d}
|T^{\rm IMP}_{Vd,\gamma d}(E)| {\bf q}\lambda,\Phi^{J_d}_{{\bf P},M_d} },
\label{eq:fsi-int}
\end{eqnarray}
where ${\bf \bar{P}}={\bf P}+{\bf q}-{\bf k}'$.

The matrix element of $T^{(1)}_{Vd,Vd}(E)$ in the above equation is defined by
Eqs.~(\ref{eq:tvnvn})-(\ref{eq:u1}). 
We follow the previous works~\cite{Kerman:1959fr,Landau:1973iv,Lee:1974zr,
Feshbach:1992} to use the FSA to calculate the first-order optical potential
$U^{(1)}_{Vd,Vd}$ and solve the scattering equation of Eq.~(\ref{eq:tvnvn}) in the
$V$-$d$ c.m frame.
Eq.~(\ref{eq:u1}) then leads to the following matrix element (omitting the spin
indices)
\begin{eqnarray}
\braket { {\bf p}|U^{(1)}_{Vd,Vd}|{\bf p}' } =
\braket{ {\bf \kappa}|t_{VN,VN}|{\bf \kappa}' } F_T(t'),
\label{eq:u1-fsa}
\end{eqnarray}
where $t'=(p'-p)^2$ and
${\bf p}'$ $({\bf p})$ is the initial (final) momentum of $J/\psi$ in the
$V$-$d$ c.m. system in which a frozen nucleon in the deuteron is in the opposite
direction with the momentum $-{\bf p}'/A_d$ ($-{\bf p}/A_d$).
We need to use the Lorentz Boost tranformation (see Eq.~(\ref{eq:smap-b}))
to evaluate ${\bf \kappa}'$
(${\bf \kappa}$), which is the momentum of $J/\psi$ in the initial (final) $V$-$N$
c.m. system, from the $(J/\psi,N)$ momenta
$({\bf p}',-{\bf p}'/A_d)$ ($({\bf p},-{\bf p}/A_d)$).  
The $J/\psi N \to J/\psi N$ scattering amplitude
$\braket{ {\bf \kappa}|t_{VN,VN}|{\bf \kappa}' }$ is obtained by solving
Eq.~(\ref{eq:eq1-c}) using the $J/\psi N$ potential generated from the quark-$N$
potential of Eq.~(\ref{eq:qn-fsi}).
$F_T(t')$ is the nuclear form factor.
For the deuteron, we use
$F_T(t')=F_{ave}(t')$ of Eq.~(\ref{eq:fave}).

\begin{figure}[t] 
\begin{center}
\includegraphics[width=0.80\columnwidth,angle=0]{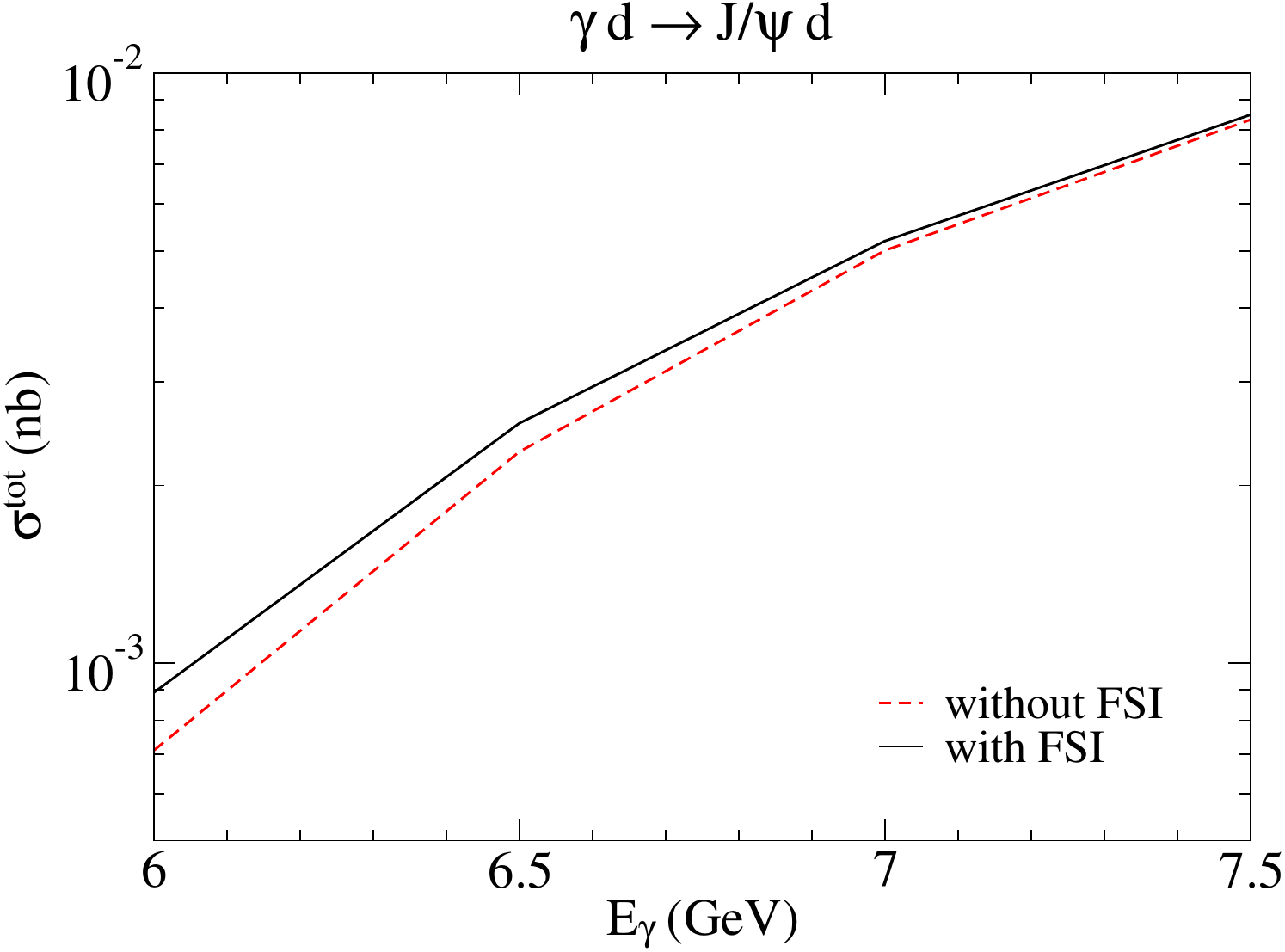}
\caption{FSI effects on the total cross section of $\gamma d \to J/\psi d$.
The results without (dashed curve) and with (solid curve) the FSI effects are
compared.}
\label{fig:fsi-on-totcrst}
\end{center}
\end{figure}

\begin{figure}[ht] 
\begin{center}
\includegraphics[width=0.80\columnwidth,angle=0]{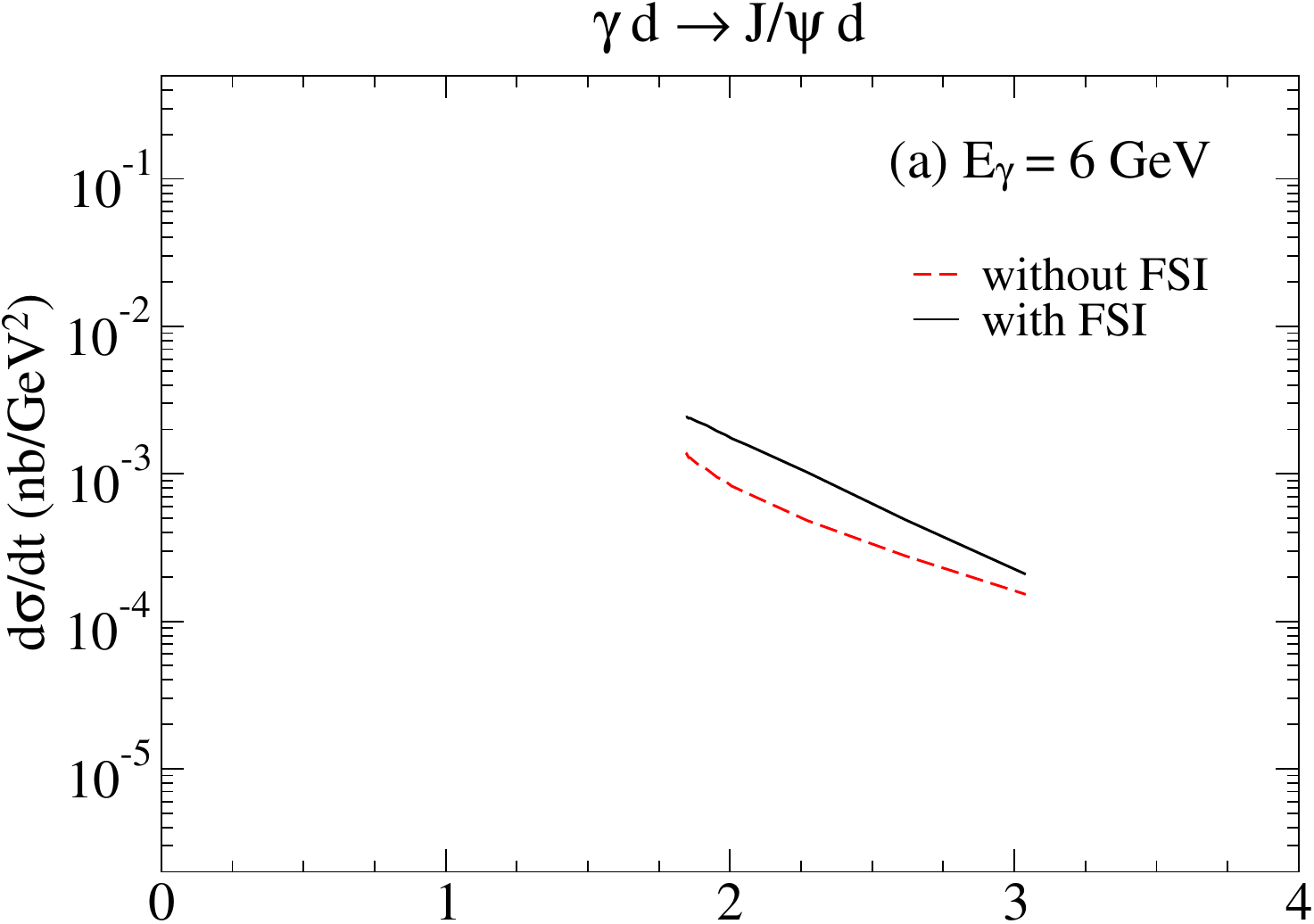} \\
\vspace{0.3em}
\includegraphics[width=0.80\columnwidth,angle=0]{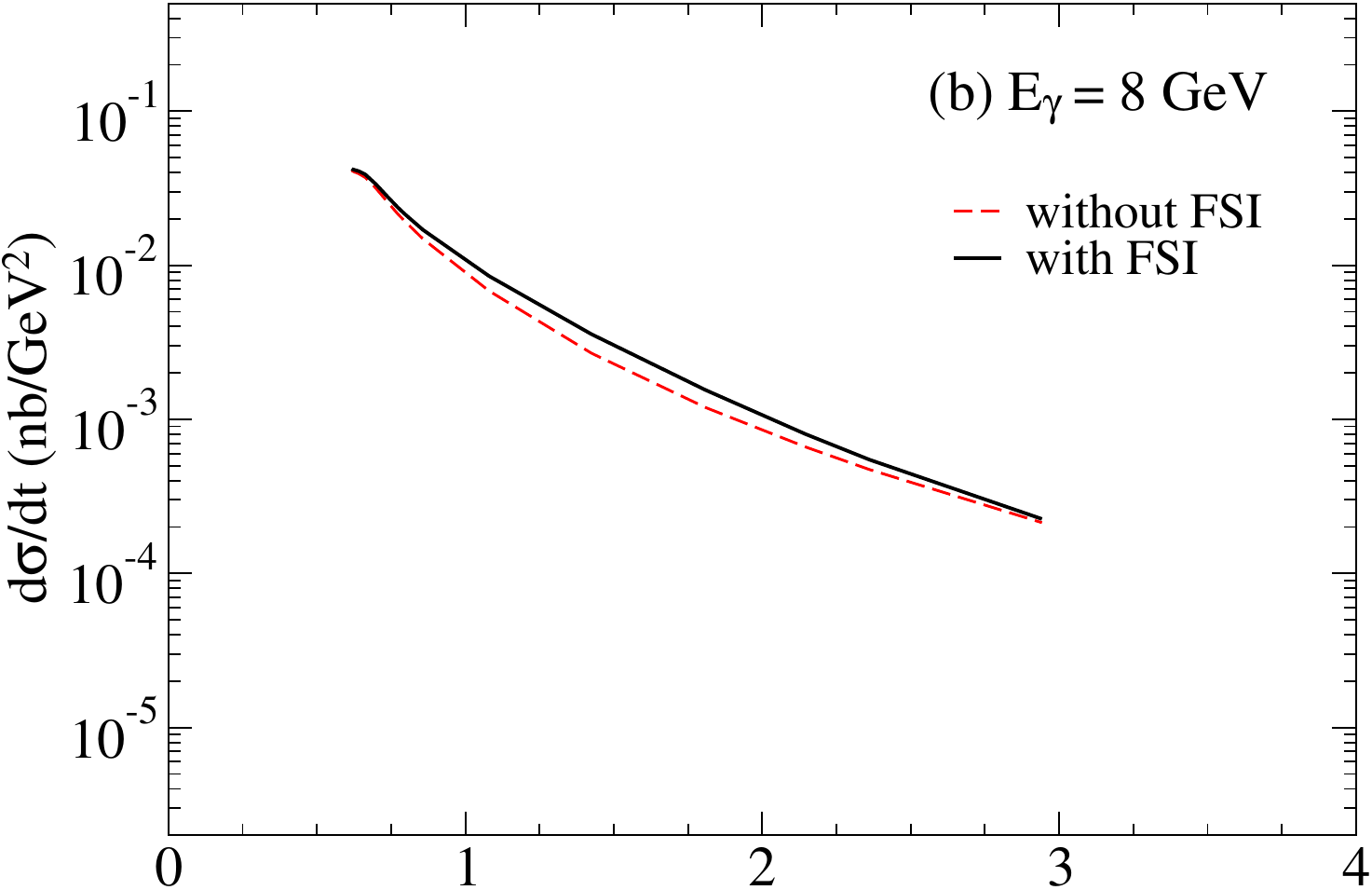} \\
\vspace{0.3em}
\includegraphics[width=0.80\columnwidth,angle=0]{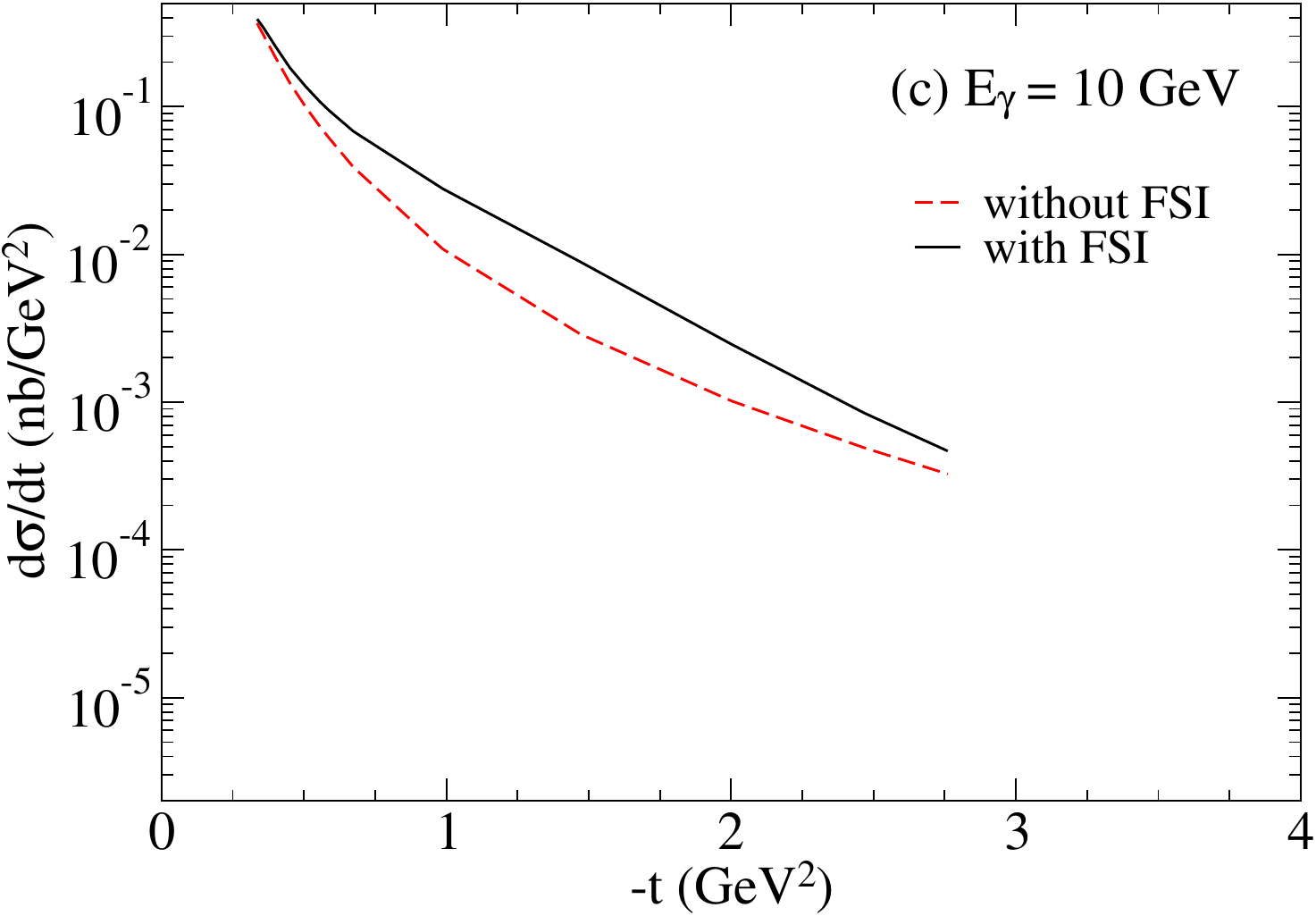}
\caption{FSI effects on the differential cross sections of $\gamma  d \to J/\psi d$
at $E_\gamma =$ (6, 8, 10) GeV.
The results without (dashed curves) and with (solid curves) the FSI effects are
compared at three different photon energies.}
\label{fig:fsi-on-dsdt}
\end{center}
\end{figure}


As given in Appendix, $\braket{ {\bf \kappa}|t_{VN,VN}|{\bf \kappa}' }$ is calculated
from the spin independent quark-$N$ potential of Eq.~(\ref{eq:qn-fsi}).
The spin averaged nuclear form factor
$F_T( t')=F_{ave}( t')$
is also independent of spin.
Thus the matrix element with spin indices of $U^{(1)}_{Vd,Vd}$ is of the form
\begin{eqnarray}
&& \Braket{ {\bf p}, m_V M_{d}|U^{(1)}_{Vd,Vd}|{\bf p}', m'_V M'_{d} }
\nonumber\\
&& = \delta_{m_V,m'_V} \delta_{M_d,M'_d}
\Braket{ {\bf p} |U_{Vd}|{\bf p}' }.
\label{eq:v-vnvn-mx}
\end{eqnarray}
Accordingly, the $J/\psi d \to J/\psi d$ scattering amplitude, defined in
Eq.~(\ref{eq:tvnvn}), can also be written as
\begin{eqnarray}
&& \Braket{ {\bf p}, m_V M_{d}|T^{(1)}_{Vd,Vd}(E)|{\bf p}', m'_V M'_{d} }
\nonumber\\
&& = \delta_{m_V,m'_V}\delta_{M_d,M'_d}
\Braket{ {\bf p} |T_{Vd}(E)|{\bf p}' },
\label{eq:t-vava-mx}
\end{eqnarray}
where $\braket{ {\bf p}|T_{Vd}(E)|{\bf p}' }$ is defined by the following
Lippmann-Schwinger equation
\begin{eqnarray}
\braket{{\bf p}|T_{Vd}(E)|{\bf p}'} &=& \braket{ {\bf p}|U_{Vd}|{\bf p}' }
\nonumber \\ &&
\hskip -1.5cm + \int d {\bf p}^{\prime\prime}
\frac{\braket{{\bf p}|U_{Vd}|{\bf p}^{\prime\prime}}
\braket{ {\bf p}^{\prime\prime} |T_{Vd}(E)|{\bf p}' } }
{E- E_V(p^{\prime\prime})-E_d (p^{\prime\prime})+i\epsilon}.
\label{eq:lseq-01}
\end{eqnarray}
We solve Eq.~(\ref{eq:lseq-01}) by using the standard numerical method described in
Ref.~\cite{Haftel:1970zz}.

Note that Eq.~(\ref{eq:fsi-int}) is given in the Lab frame.
Following Ref.~\cite{Keister:1991sb}, the $J/\psi$-$d$ scattering amplitude in
Eq.~(\ref{eq:fsi-int}) is then calulated from Eq.~(\ref{eq:t-vava-mx}) by 
\begin{eqnarray}
&&\braket{ {\bf k}m_V,\Phi^{J_d}_{{\bf P}^\prime,M^\prime_d}|T^{(1)}_{Vd,Vd}(E)|
{\bf k}^\prime \bar{m}_V,\Phi^{J_d}_{{\bf \bar{P}},\bar{M}_d} } \nonumber \\
&&= \delta_{m_V,\bar{m}_V} \delta_{M'_d,\bar{M}_d}
\braket{{\bf k} {\bf P}' | T_{Vd}(E) | {\bf k}' \bar{\bf P}},
\end{eqnarray}
where 
\begin{eqnarray}
&&\braket{{\bf k} {\bf P}' | T_{Vd}(E) |  {\bf k}' \bar{\bf P}} \nonumber \\
&&=N({\bf p}',{\bf k} {\bf P}')\braket{ {\bf p}' | T_{Vd}(E) | \bar{\bf p} }
 N(\bar{{\bf p}},{\bf k}' \bar{{\bf P}}),
\label{eq:re-tvd}
\end{eqnarray}
with
\begin{eqnarray}
\hskip -0.5cm
N({\bf p}, {\bf p}_1 {\bf p}_2) =
\left[ \frac{E_1({\bf p}_1)+E_2({\bf p}_2)}
{E_1({\bf p})+E_2({\bf p})}\frac{E_1({\bf p})E_2({\bf p})}
{E_1({\bf p}_1)E_2({\bf p}_2)} \right]^{\frac12},\,\,\,
\end{eqnarray}
where ${\bf p}$ is the momentum in the $V$-$d$ c.m. frame calculated from the momenta
${\bf p}_1$ and ${\bf p}_2$ in the Lab frame.
To be consistent with the procedures for calculating  $T^{(1)}_{Vd,Vd}(E)$, we use the
FSA to evaluate the matrix element of $T^{\rm IMP}_{V d,\gamma d} (E)$ in the FSI
amplitude of Eq.~(\ref{eq:fsi-int}).
Namely, $T^{\rm IMP}_{Vd,\gamma d} (E)$ in Eq.~(\ref{eq:fsi-int}) is calculated from
Eq.~(\ref{eq:gd-amp-0}) by setting the initial nucleon momentum to be
${\bf p}=0$.

\begin{figure}[t] 
\begin{center}
\includegraphics[width=0.80\columnwidth,angle=0]{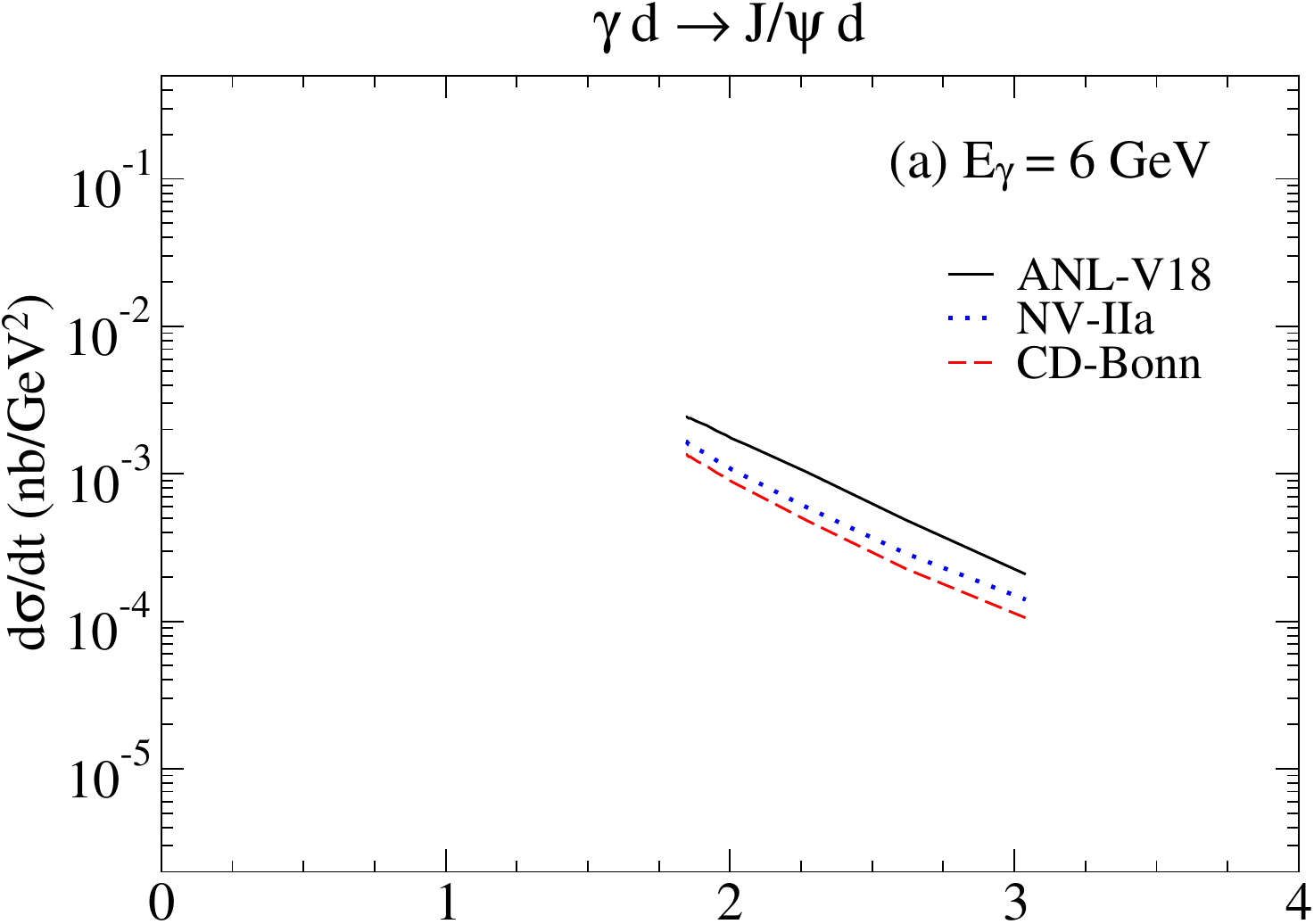} \\
\vspace{0.5em}
\includegraphics[width=0.80\columnwidth,angle=0]{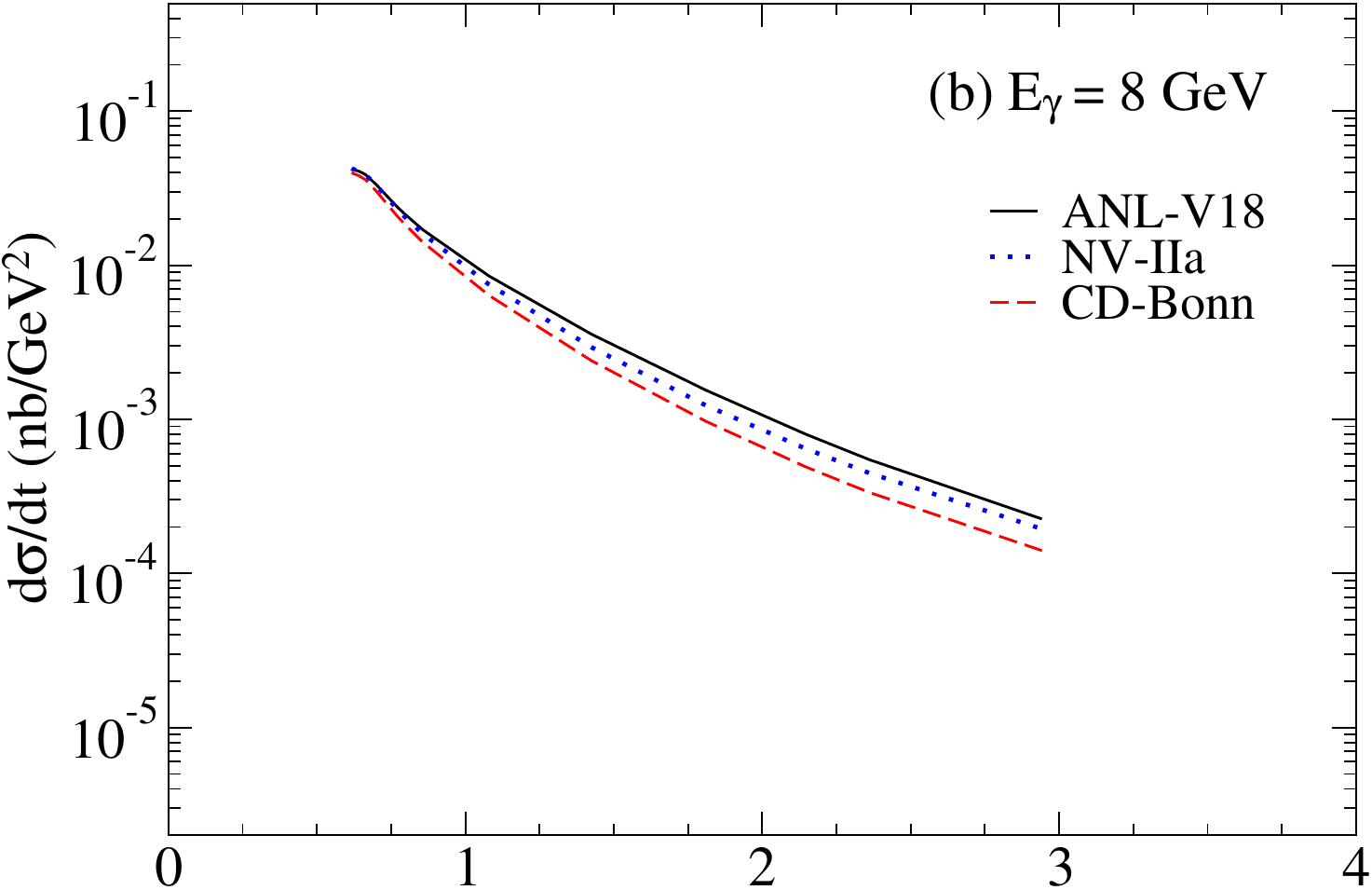} \\
\vspace{0.5em}
\includegraphics[width=0.80\columnwidth,angle=0]{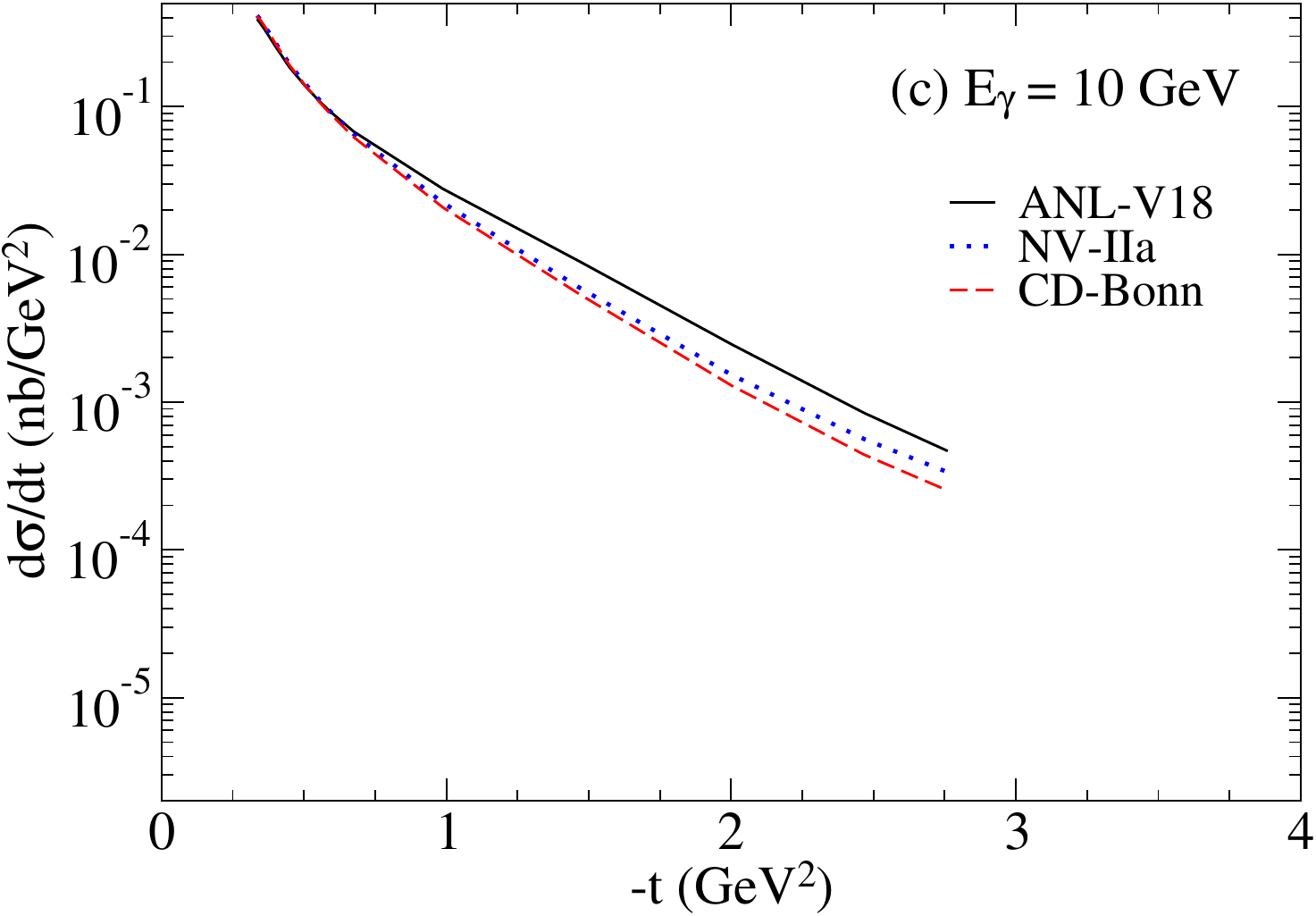}
\caption{Differential cross sections of $\gamma d \to J/\psi d$ at $E_\gamma =$
(6, 8, 10) GeV.
The results using the deuteron wave function generated from three different $NN$
potentials~\cite{Wiringa:1994wb,Piarulli:2016vel,Machleidt:1989tm} are compared.}
\label{fig:gd-comp-dsdt-pot}
\end{center}
\end{figure}

The FSI effects are presented in Fig.~\ref{fig:fsi-on-totcrst} where we compare the
total cross section of $T^{\rm IMP}_{Vd,\gamma d}$ (dashed curve) and
$T^{\rm IMP}_{Vd,\gamma d} + T^{\rm FSI}_{Vd,\gamma d}$ (solid curve).
Here $T^{\rm IMP}_{Vd,\gamma d}$ indicates the result obtained from the relativistic
formulation of Eq.~(\ref{eq:gd-amp-0}) explained in Sec.~\ref{Section:IV}(A).
We see that the FSI effects significantly increase the cross sections at energies
near the threshold.
This can be understood from the results of the differential cross sections shown in
Fig.~\ref{fig:fsi-on-dsdt}.
We find that the FSI amplitude mainly contributes to the large $-t$ region.  

For later calculations of the $J/\psi$ photo-production on heavier nuclei in
Sec.~\ref{Section:V}, we note here that, if we use Eq.~(\ref{eq:t-fsa}) of the FSA to
evaluate $T^{\rm IMP}$ of Eq.~(\ref{eq:fsi-int}), Eq.~(\ref{eq:fsi-int}) can then be
simplified as 
\begin{eqnarray}
&& \braket{ {\bf k} m_V, \Phi^{J_d}_{{\bf P}^\prime,M^\prime_d} | T^{\rm FSI}_{Vd,\gamma d} (E)
|{\bf q} \lambda, \Phi^{J_d}_{{\bf P}, M_d} } 
\nonumber \\
&& \sim A_d \braket{ {\bf k} m_V, {\bf t} \bar{m}_{s'_1} | \bar{t}_{VN,\gamma N} (\omega_0)
|{\bf q} \lambda, {\bf 0} \bar{m}_{s_1} }
\nonumber \\
&& \times F_{T}^{\rm FSI}(t) \delta_{M'_d,M_d},
\label{eq:amp-fact-d-fsi}
\end{eqnarray}
where
\begin{eqnarray}
F_T^{\rm FSI}(t) &=&
\int d{\bf k}^\prime
\braket{ {\bf k}{\bf P}'|T_{Vd}(E)| {\bf k}^\prime \bar{{\bf P}} }
\nonumber \\
&& \times \frac{1}{E-E_V(k')-E_d(\bar{P})+i\epsilon} F_{ave}(t).
\label{eq:f-fsi-1}
\end{eqnarray}
By using Eqs.~(\ref{eq:amp-fact-d-fsi})-(\ref{eq:f-fsi-1}),
the FSA cross section of Eq.~(\ref{eq:cross-fsa}) can then be extended to include
the FSI term and becomes
\begin{eqnarray}
\left( \frac{d\sigma}{d\Omega_{\rm Lab}} \right)_{\rm FSA} &=&
\rho_d({\bf k},{\bf q})
\frac{1}{4} \sum_{m_{s'_1},m_{s_1},m_V,\lambda}
\nonumber \\
&& \hskip -1.8cm
\times | \braket { {\bf k}m_V,{\bf t}{m}_{s'_1}|{t}_{VN,\gamma N}(\omega)
|{\bf q}\lambda,{\bf 0}{m}_{s_1} } |^2
\nonumber \\
&& \hskip -1.8cm
\times | A_d [F_{ave}( t)+F_T^{\rm FSI}( t')] |^2.
\label{eq:cross-fsa-tot}
\end{eqnarray}

\subsection{Model dependence}

\begin{figure}[t] 
\begin{center}
\includegraphics[width=0.80\columnwidth,angle=0]{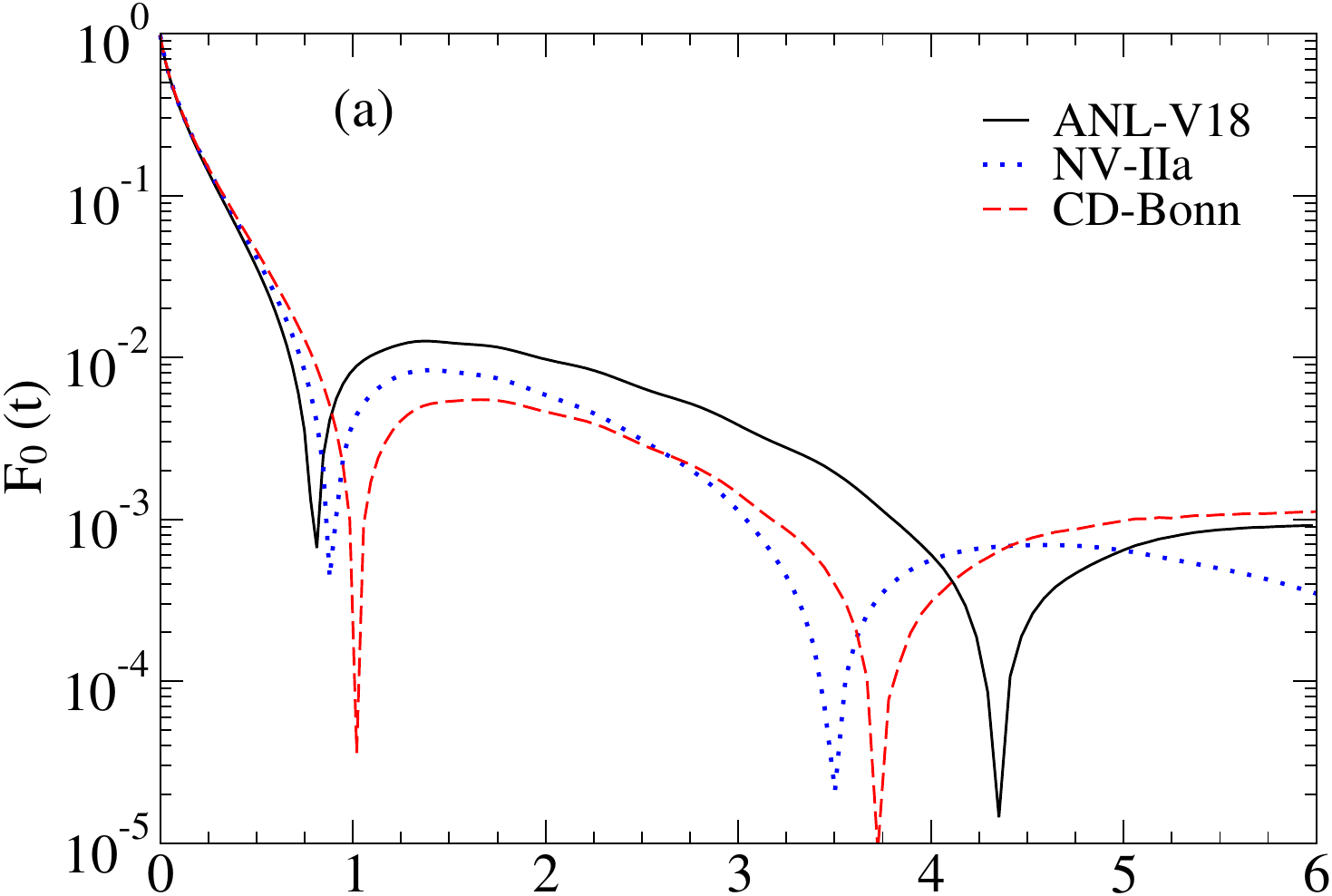} \\
\vspace{0.5em}
\includegraphics[width=0.80\columnwidth,angle=0]{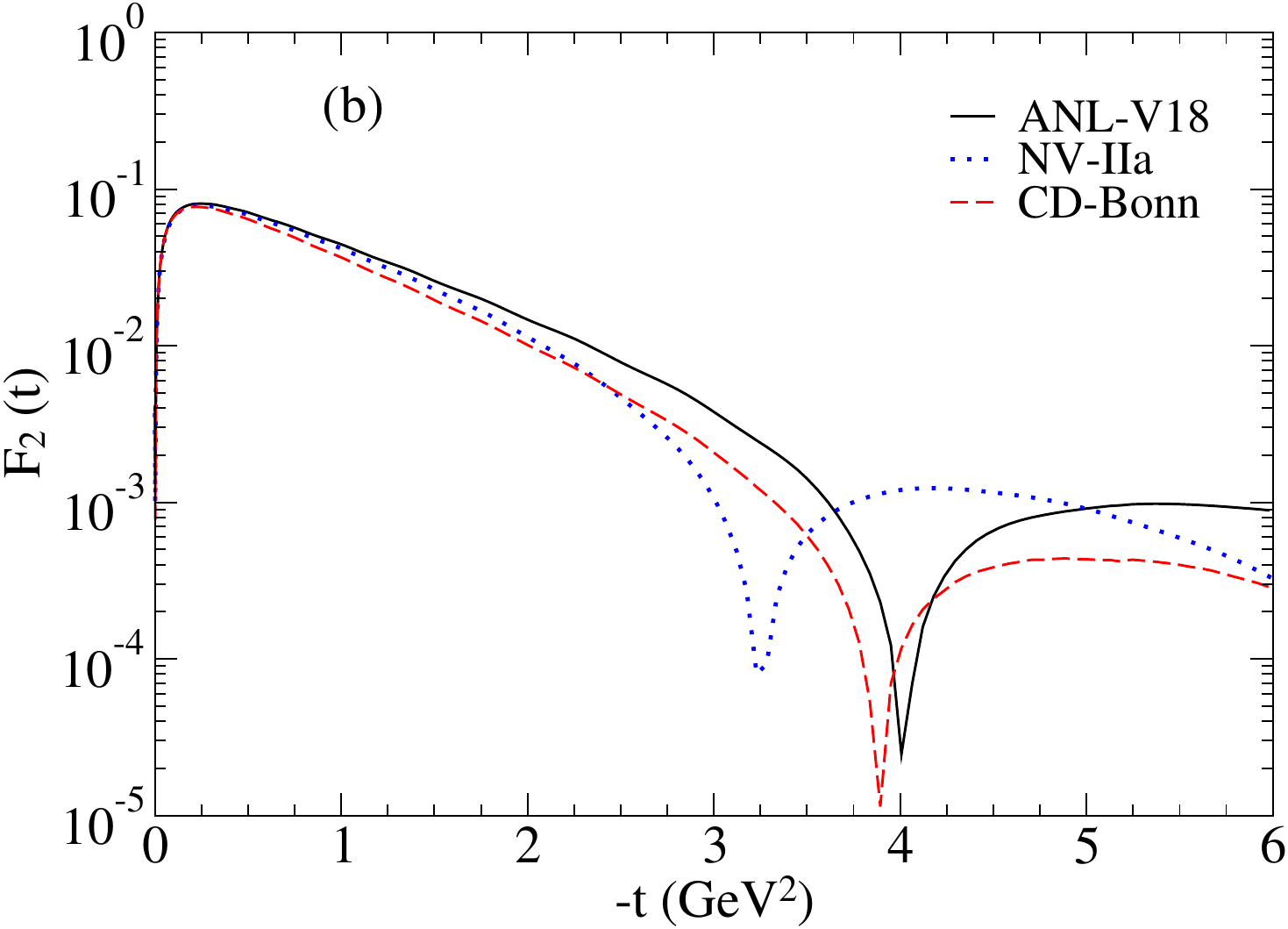}
\caption{Deuteron form factors (a) $F_0(t)$ and (b) $F_2(t)$
calculated from the deuteron wave function gennerated from three different $NN$
potentials~\cite{Wiringa:1994wb,Piarulli:2016vel,Machleidt:1989tm} are compared.}
\label{fig:gd-comp-ff-pot}
\end{center}
\end{figure}

\begin{figure}[t] 
\begin{center}
\includegraphics[width=0.85\columnwidth,angle=0]{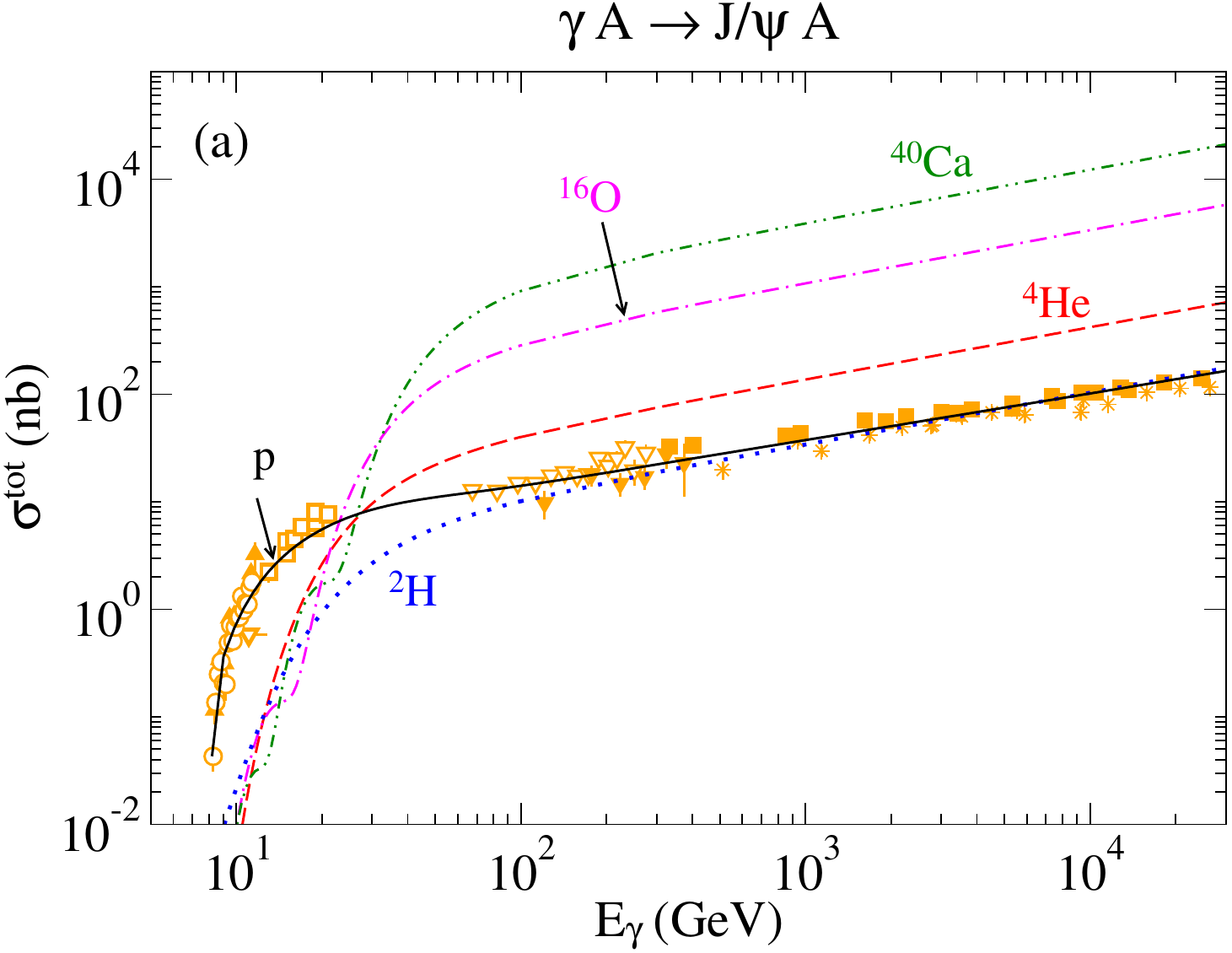} \\
\vspace{0.7em}
\includegraphics[width=0.85\columnwidth,angle=0]{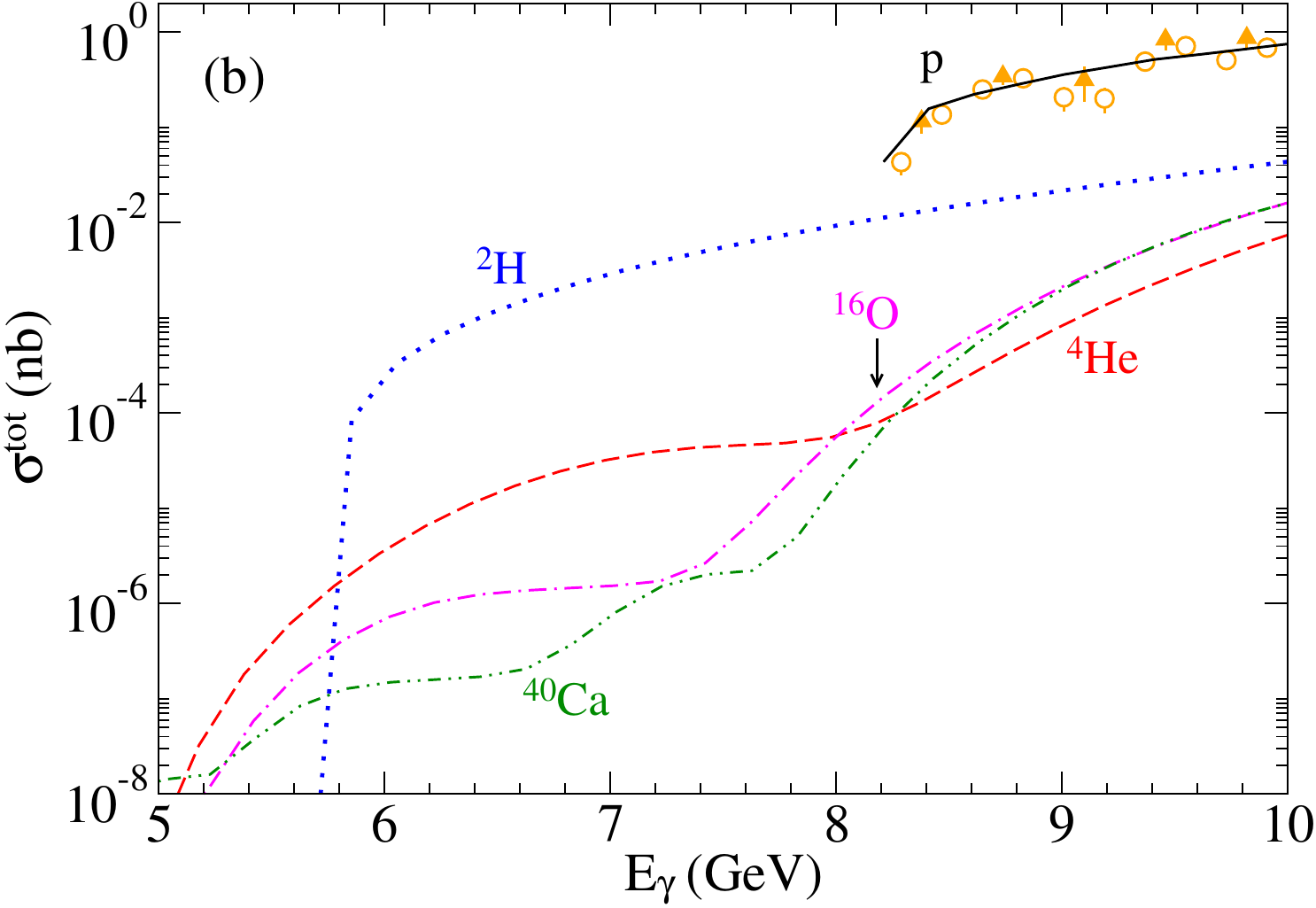}
\caption{(a) Total cross sections of $\gamma A \to J/\psi A$ ($A =$ $p$, $^2{\rm H}$,
${^4\rm He}$, ${^{16}\rm O}$, and ${^{40}\rm Ca}$) are plotted as a function of
$E_\gamma$ from threshold up to $E_\gamma \sim 10^4$ GeV.
(b) Total cross section at low energies ($5 \leqslant E_\gamma \leqslant 10$ GeV).
The SLAC~\cite{Camerini:1975cy}, FermiLab~\cite{Binkley:1981kv,E687:1993hlm},
ZEUS~\cite{ZEUS:2002wfj}, H1~\cite{H1:1996kyo,H1:2000kis}, and
GlueX~\cite{GlueX:2019mkq,GlueX:2023pev} data are used for $\gamma p \to J/\psi p$.}
\label{fig:TCS-all}
\end{center}
\end{figure}

By using the calculation procedures described in Sec.~\ref{Section:IV}(C), we next
investigate how the predicted cross sections depend on the $NN$ model used in
generating the deuteron wave function.
We consider the Argonne-V18~\cite{Wiringa:1994wb}, Norfolk
NV-IIa~\cite{Piarulli:2016vel}, and CD-Bonn~\cite{Machleidt:1989tm} models.
The predicted differential cross sections are compared in
Fig.~\ref{fig:gd-comp-dsdt-pot}.
Here the solid curves correspond to the solid curves in Fig.~\ref{fig:fsi-on-dsdt}.
Clearly, their differences are mainly in the large $-t$ region.
This can be understood from comparing their form factors, shown in
Fig.~\ref{fig:gd-comp-ff-pot}, which are related to their wave function.
We see that their differences are also in the large $-t$ region.
Obviously, their differences can be clealy distinguished at very near threshold
$E_\gamma=6$ GeV, while it will be very challenging to measure such small cross
sections.

\section{Production on $J=0$ nuclei}
\label{Section:V}

\begin{figure*}[ht] 
\begin{center}
\includegraphics[width=1.8\columnwidth,angle=0]{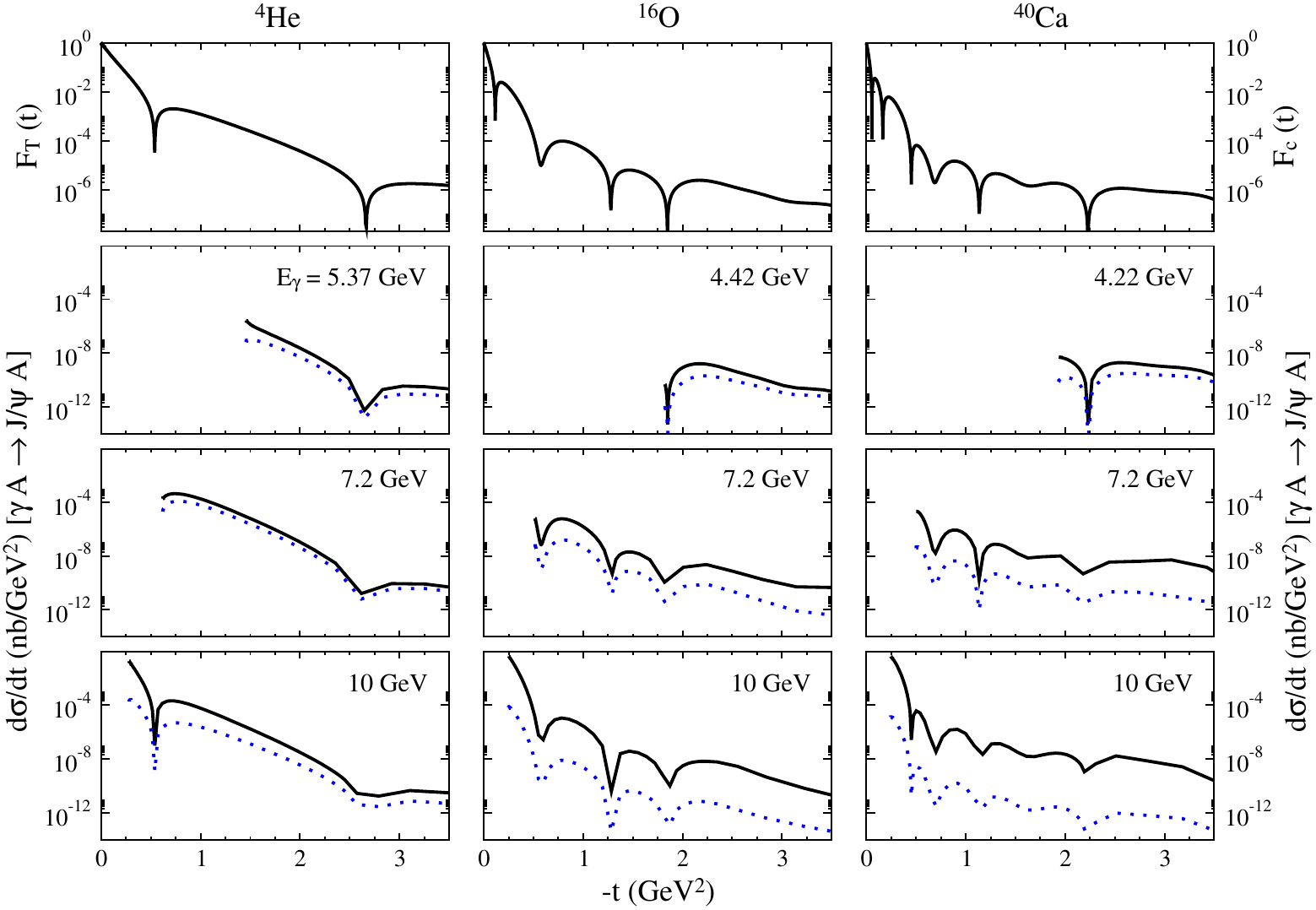}
\caption{The top panels show the nuclear form factors of nuclei $A$ ($A = {^4\rm He}$,
${^{16}\rm O}$, ${^{40}\rm Ca}$).
Below the top panels, Differential cross sections of $\gamma A \to J/\psi A$  are
presented at $E_\gamma =$ ($E_\gamma^{\rm th} + 1.0$, 7.2, 10) GeV.
The results of the FSI effects (dotted curves) and the full results (solid curves)
are compared.}
\label{fig:dsdt_A}
\end{center}
\end{figure*}

For heavier nuclei, we will use the FSA as developed in the previsous section.
We only consider the exclusive $J/\psi$ photo-production on the spin $J=0$ nuclei
which have only one nuclear form factor $F_T(t)$.
Following Eq.~(\ref{eq:cross-fsa-tot}), the differential cross sections for
$\gamma({\bf q}) + A \to J/\psi({\bf k})+ A$ in the Lab frame can be written as
\begin{eqnarray}
\frac{d\sigma}{d\Omega_{\rm Lab}} &=&
\frac{(2\pi)^4 \left\vert \textbf{k} \right\vert^2
E_{V}(\textbf{k}) E_A(\textbf{q}-\textbf{k})}
{\left\vert E_A(\textbf{q} - \textbf{k}) \vert \textbf{k} \vert + E_{V}(\mathbf{k})
( \vert \mathbf{k} \vert - \vert \bm{q} \vert \cos\theta_{\rm k}) \right\vert}
\nonumber \\ &&
\hskip -0.5cm \times \frac{1}{4} \sum_{m_{s'_1},m_{s_1},m_V,\lambda}
\nonumber \\ &&
\hskip -0.5cm \times |
\braket{ {\bf k}m_V,{\bf t}{m}_{s'_1}|{t}_{VN,\gamma N}( \omega)
|{\bf q}\lambda,{\bf 0}{m}_{s_1} }|^2
\nonumber \\
&& \hskip -0.5cm \times | A_T
[F_{T} ( t) + F^{\rm FSI}_T ( t')] |^2,
\label{eq:cross-fsa-tot-t}
\end{eqnarray}
where
$t=(q-k)^2$ and $t'=(k'-k)^2$.
$A_T$ is the number of nucleons in the target nucleus $T$ and
\begin{eqnarray}
F^{\rm FSI}_T (t') &=&
\int d{\bf k}^\prime
\braket{ {\bf k}{\bf P}'|T_{VA}(E)| {\bf k}^\prime \bar{{\bf P}} }
\nonumber \\
&& \times \frac{1}{E-E_V(k')-E_A(\bar{P})+i\epsilon} F_T(t'),
\label{eq:f-fsi-2}
\end{eqnarray}
with  $\bar{{\bf P}}={\bf q}-{\bf k}'$ and ${\bf P}'= {\bf q}-{\bf k}$.
The form factor $F_T(t)$ in Eq.~(\ref{eq:cross-fsa-tot-t}) and (\ref{eq:f-fsi-2})
is calculated from~\cite{Lonardoni:2017egu}
\begin{eqnarray}
F_T(t)= \braket{ \Psi_T \vert \sum_{i=1,A}e^{i \bm{\kappa} \cdot \bm{r}^{}_i} \vert \Psi_T},
\label{eq:ft}
\end{eqnarray}
where $\ket{\Psi_T}$, normalized as $\braket{\Psi_T \vert \Psi_T} = 1$,
is the nuclear ground state.
$\bm{\kappa}$ is the three-momentum transfer to the nucleus in the rest frame
of the nucleus and is related to $t$ by
\begin{eqnarray}
-t = \bm{\kappa}^{2}-\omega^2,
\end{eqnarray}
with
\begin{eqnarray}
\omega =\sqrt{\bm{\kappa}^{\,\,2}+m^2_T}-m_T^{},
\end{eqnarray}
where $m_T^{}$ is the mass of the target nucleus $T$.
Clearly, $F_T(t)$ is related to the nuclear charge form factor $F_c(q^2)$
(with no exchange current contribution) by
\begin{eqnarray}
F_c(q^2)=F_N(q^2)F_T(q^2=t),
\end{eqnarray}
where $F_N(q^2)$ is the nucleon charge form factor.

The above formulae are the same as those employed in Ref.~\cite{Kim:2021adl}
to investigate the $\phi$ photo-production on ${^4\rm He}$.
In this work, the nuclear form factors are calculated from the wave function generated
from the Variational Monte Carlo (VMC) calculations~\cite{Lonardoni:2017egu} for
${^4\rm He}$, ${^{16}\rm O}$, and ${^{40}\rm Ca}$. 
The predicted total cross sections are compared with that from the proton target in
Fig.~\ref{fig:TCS-all}.
At higher energies, the total cross sections are larger when heavier nuclei targets
are used as displayed in Fig.~\ref{fig:TCS-all}(a).
Meanwhile, the total cross sections near the threshold are more dramatic as seen in
Fig.~\ref{fig:TCS-all}(b).
They are very small compared with that from the proton target (solid curve).
This is not surprising, because near the threshold the cross sections are mainly
from large $-t$ region where the nuclear form factors are small, as seen clealy in
the top panels of Fig.~\ref{fig:dsdt_A} where the predicted differential cross
sections are also presented.
Here we compare the results of $T^{\rm FSI}$ and $T^{\rm IMP} + T^{\rm FSI}$.
We find that the FSI effects are significant near the threshold but
are very weak at rather higher energies.
The measurements for heavier nuclei targets near the threshold will be very challenging
and help shed light on the relevant reaction mechanism.

\section{Summary and Future developments}
\label{Section:VI}

We have used the Pom-CQM model~\cite{Lee:2022ymp,Sakinah:2024cza} for the $\gamma p
\to J/\psi p$ reaction to predict the cross sections of the exclusive $J/\psi$
photo-production on nuclei. 
The calculations have been performed within the multiple scattering theory by
including the impulse amplitude $T^{\rm IMP}_{J/\psi A,\gamma A}$ and the final
$J/\psi$-nucleus scattering amplitude $T^{\rm FSI}_{J/\psi A,\gamma A}$.
The impulse term $T^{\rm IMP}_{J/\psi A,\gamma A}$ for the deuteron target can be
calculated exactly using the wave function generated from the realistic
$NN$ potentials.
We have shown that, near the threshold region, the $J/\psi$ photo-production cross
sections depend strongly on the $d$-state of the deuteron wave function. 
We also find that the conventional FSA, which expresses the amplitude
$T^{\rm IMP}_{J/\psi A,\gamma A}$ in terms of the averaged $\gamma p \to J/\psi p$
amplitude and the nuclear form factor, is not valid near the threshold region
but is a good approximation at higher energies.

The FSA is then applied to predict the cross sections of the $J/\psi$ photo-production
on the ${^4\rm He}$, ${^{16}\rm O}$, and ${^{40}\rm Ca}$ targets using their nuclear
form factors from the variatioal Monte-Carlo calculations of
Ref.~\cite{Lonardoni:2017egu}.

The $J/\psi$-nucleus scattering amplitude, which is needed to evaluate the
FSI amplitude $T^{\rm FSI}_{J/\psi A,\gamma A}$, is calculated using the first-order
optical potentials which are constructed using the nuclear form factor and the
$J/\psi$-$N$ scattering amplitude generated from the employed Pom-CQM model.
It is found that the FSI effects are significant at large $-t$ region.

The Pom-CQM model used in this work is based on the phenomenological quark-$N$
potentials $v_{cN}(r)$ determined by fitting to the JLab data.
To interprete our results in terms of the role of gluons in nuclei, it is necessary
to relate $v_{cN}(r)$ to the gluons associated with the nucleon.
If we also use the CQM model for nucleon, $v_{cN}(r)$ can be derived by folding a
quark-quark interaction potential due to the gluon-exchange mechanisms into the
constituent quark density of the nucleon.
If we use the partonic model for the nucleon, a possibility is to relate $v_{cN}(r)$
to the GPD-based model of $J/\psi$ photo-production model of Ref.~\cite{Guo:2021ibg}.
Either one is non-trivial but must be pursued in the future.

Our predictions for $A > 2$ nuclei near the threshold have been done only using
the FSA, and thus are not reliable in the sub-threshold region where the bound nuclei
with hidden charms~\cite{Brodsky:1989jd,Gao:2000az,Belyaev:2006vn,Wu:2012wta} or
$J/\psi$-nucleus resonances, if exist, can be identified.
To make progress in this direction, it is necessary to go beyond the FSA and perform
the calculation using the nuclear many-body wave function.

\section*{Appendix}
\label{Appendix:A}

The formulation of the Pom-CQM model was given in detail in
Refs.~\cite{Lee:2022ymp,Sakinah:2024cza}.
Here we only give the formulae which are used for the calculations in this work and
are needed to develop the formulation for the $J/\psi$ photo-production on nuclei.

In the c.m. frame, the matrix element of $T^D_{VN,\gamma  N}$ of Eq.~(\ref{eq:eq1-a})
can be decomposed into
\begin{eqnarray}
&& \Braket{ {\bf p}', m_V m_{s'} | T^D_{VN,\gamma N}(W) | {\bf q}, \lambda m_s }
\nonumber  \\
&&=\Braket{{\bf p}', m_V m_{s'} | B_{VN,\gamma N}(W) | {\bf q}, \lambda m_s}
\nonumber \\
&&+ \braket{ {\bf p}', m_V m_{s'} | T^{\rm FSI}_{VN,\gamma N} (W) | {\bf q},\lambda m_s },
\label{eq:pom-cqm-t0}
\end{eqnarray}
where $W$ is the invariant mass in the initial $\gamma$-$N$ system.
The Born term illustrated in Fig.~\ref{fig:gnvn} can be expressed as
\begin{eqnarray}
&& \Braket{ {\bf p}', m_V m_{s'} | B_{VN,\gamma N}(W) | {\bf q}, \lambda m_{s} } 
\nonumber\\
&& = C_{\lambda,m_V} \delta_{m_s,m_{s'}} B({\bf p}',{\bf q},W)
\left[ 2 v^{B}_{cN}({\bf q}-{\bf p}') \right] ,
\label{eq:mtx-b-1}
\end{eqnarray}
with 
\begin{eqnarray}
C_{\lambda,m_V} = \sum_{m_c,m_{\bar{c}}} \textstyle
\braket{ J_Vm_V | \frac{1}{2} \frac{1}{2} m_c m_{\bar c} }
\braket{ m_{\bar{c}}|{\bf \sigma} \cdot {\bf \epsilon_\lambda}|m_c },
\end{eqnarray}
and
\begin{eqnarray}
v^B_{cN}({\bf q}-{\bf p}') = \frac{1}{(2\pi)^3}\int\, d{\bf r}\,
e^{i{\bf ({\bf q}-{\bf p}')}\cdot {\bf r}} v^B_{cN}(r).
\label{eq:vcn-t}
\end{eqnarray}
We note here that, for a Yukawa form of Eq.~(\ref{eq:qn-B}) for $v^B_{cN}(r)$, 
Eq.~(\ref{eq:vcn-t}) leads to
\begin{eqnarray}
&& \hskip -0.5cm v^B_{cN}({\bf q}-{\bf p}') \nonumber \\
&& \hskip -0.5cm = \frac{\alpha}{(2\pi)^2}
\left[ \frac{1}{({\bf q}-{\bf p}')^2+\mu_B^2} -
\frac{1}{({\bf q}-{\bf p}')^2+\mu_{B}^{'2}} \right].
\label{eq:vcn-y-t}
\end{eqnarray}
The term $B({\bf p}',{\bf q},W)$ in Eq.~(\ref{eq:mtx-b-1}) involves a loop-integration
over the $J/\psi$ wave function $\phi(k)$ and is of the following form
\begin{eqnarray}
B({\bf p}',{\bf q},W)
&=& \frac{1}{(2\pi)^3}\frac{e_c}{\sqrt{2|{\bf q}|}}
\int d{\bf k} \, \phi \left({\bf k}-{\textstyle\frac{1}{2}}{\bf p}' \right) 
\nonumber\\
&& \hskip -1.8cm \times
\frac{1}{W-E_N({\bf q})-E_c({\bf q-k})-E_c({\bf k})+i\epsilon} 
\nonumber  \\ 
&& \hskip -1.8cm \times
\sqrt{\frac{E_c({\bf k})+m_c}{2E_c({\bf k})}}
\sqrt{\frac{E_c({\bf q- k})+m_c}{2E_c({\bf q- k})}}
\nonumber\\
&& \hskip -1.8cm \times
\left\{ 1-\frac{{\bf k}\cdot{\bf (q-k)}}{[E_c({\bf k})+m_c][E_c({\bf q- k})+m_c]}
\right\},
\label{eq:tmx-b-2}
\end{eqnarray}    
where $m_c$ is the constituent quark mass of charm quark $c$,
$E_a({\bf p}) = \sqrt{m_a^2+{\bf p}^2}$ the energy for the particle $a$ with mass
$m_a$, and $e_c=\frac{2}{3}e$ with $e$ being the electron charge.

\begin{figure}[h]
\centering
\includegraphics[width=0.8\columnwidth,angle=0]{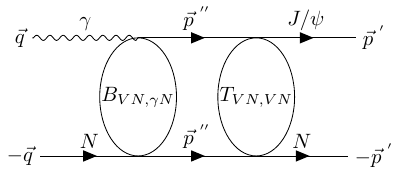}
\caption{$J/\psi$ photo-production on the nucleon with the FSI effects.}
\label{fig:tfsi-mx}
\end{figure}

The FSI term in Eq.~(\ref{eq:pom-cqm-t0}) is illustrated in Fig.~\ref{fig:tfsi-mx}
and is of the form
\begin{eqnarray}
&& \Braket{ {\bf p}',m_Vm_{s'}|T^{\rm FSI}_{VN,\gamma N}(W)|{\bf q},\lambda m_s }
\nonumber\\ &=&
\sum_{m^{\prime\prime}_V,m^{\prime\prime}_s} \int d{\bf p}^{\prime\prime} 
\Braket{ {\bf p}', m_V m_{s'} |T_{VN,VN}(W)|
{\bf p}^{\prime\prime}, m^{\prime\prime}_V m^{\prime\prime}_s } 
\nonumber \\ && \times
\frac{1}{W-E_V(p^{\prime\prime})-E_N(p^{\prime\prime})+i\epsilon}
\nonumber\\ && \times
\Braket{{\bf p}^{\prime\prime}, m^{\prime\prime}_V m^{\prime\prime}_s
|B_{VN,\gamma N}(W)|{\bf q}, \lambda m_s}. 
\label{eq:tfsi-mx}
\end{eqnarray}
With the spin independent quark-$N$ potential defined by Eq.~(\ref{eq:qn-fsi}),
the $J/\psi N \to J/\psi N$ scattering amplitude in the above equation can be written
as
\begin{eqnarray}
&& \Braket{ {\bf p}', m'_V m_{s}'|T_{VN,VN}(W)|{\bf p}, m_V m_{s} }
\nonumber\\
&& = \delta_{m_V,m'_V}\delta_{m_s,m'_s}
\Braket{ {\bf p}' |T_{VN}(W)|{\bf p} },
\label{eq:t-vnvn-0}
\end{eqnarray}
where $\braket{ {\bf p}'|T_{VN}(W)|{\bf p} }$ is defined  by the following
Lippmann-Schwinger equation
\begin{eqnarray}
\braket{{\bf p}'|T_{VN}(W)|{\bf p}} &=& \braket{ {\bf p}'|V_{VN}|{\bf p} } 
\nonumber \\
&& \hskip -1.8cm
+ \int d{\bf p^{\prime\prime}}
\frac{\braket{{\bf p}'|V_{VN}|{\bf p}^{\prime\prime}}\braket{{\bf p}^{\prime\prime}
|T_{VN}(W)|{\bf p}}}
{W- E_V(p^{\prime\prime})-E_N(p^{\prime\prime})+i\epsilon}.
\label{eq:lseq-00}
\end{eqnarray}
Here $\braket{ {\bf p}'|V_{VN}|{\bf p} }$, illustrated in Fig.~\ref{fig:vnvn}, is
obtained by folding $v_{cN} ^{\rm FSI}(r)$ of Eq.~(\ref{eq:qn-fsi}) into the $J/\psi$
wave function and is of the following form
\begin{eqnarray}
\braket{ {\bf p}'|V_{VN}|{\bf p} }
= F_V({\bf t}) \left[ 2v^{\rm FSI}_{cN}({\bf t}) \right] ,
\end{eqnarray}
where ${\bf t} = {\bf p}-{\bf p}'$ and
\begin{eqnarray}
v^{\rm FSI}_{cN}({\bf q}-{\bf p}')
&=& \frac{1}{(2\pi)^3}\int\, d{\bf r}\,
e^{i{\bf ({\bf q}-{\bf p}')}\cdot {\bf r}} v^{\rm FSI}_{cN}(r),
\label{eq:vfsi} \\
F_V({\bf t}) &=& \int  d{\bf k} \, \textstyle
\phi^*\left({\bf k}-\frac{{\bf t}}{2}\right) \phi\left({\bf k}\right).
\label{eq:v-ff}
\end{eqnarray}
Here $F_V({\bf t})$ is the form factor of the vector meson $V$ and $\phi({\bf k})$ 
is the wave function of $J/\psi$  in momentum space.
For a Yukawa form of $v_{cN}^{\rm FSI}(r)$  of Eq.~(\ref{eq:qn-fsi}),
Eq.~(\ref{eq:vfsi}) has the same form as Eq.~(\ref{eq:vcn-y-t}).

Following the approach of Donnachie and Landshoff~\cite{Donnachie:1984xq,
Donnachie:1992ny,Donnachie:1994zb,Donnachie:1998gm}, the Pomeron-exchange amplitude
$T^{\rm Pom}_{VN,\gamma N}(W)$ in Eq.~(\ref{eq:t-0}) is constructed within the Regge
phenomenology and is of the form
\begin{eqnarray}
&&\braket{ \mathbf{k},m_V m_{s'} | T^{\rm Pom}_{VN,\gamma N}(W) | \mathbf{q},\lambda m_s }
\nonumber\\
&&=\frac{1}{(2\pi)^3} \sqrt{
\frac{m_N m_N}{4 E_V(\mathbf{k}) E_N(\mathbf{p}')|\mathbf{q}|E_N(\mathbf{p})} }
\nonumber \\ && \times
[ \bar{u}(p',m_{s'}) \epsilon^*_\mu(k,m_{V}) \mathcal{M}^{\mu\nu}_{\mathbb P}(k,p';q,p)
\nonumber\\ && \times
\epsilon_\nu(q,\lambda) u(p,m_s) ],
\end{eqnarray}
where 
\begin{eqnarray}
\mathcal{M}^{\mu\nu}_{\mathbb P}(k,p';q,p) &=&
i \frac{2 e \, m_V^2}{f_V}
( \slashed{q} g^{\mu\nu} - q^\mu \gamma^\nu )
\nonumber\\
&& \hskip -1.3cm \times [2\beta_{q_{V}}F'_V(t)]G_{\mathbb P}(s,t)[3\beta_{u/d} F_1(t)].
\label{eq:MP}
\end{eqnarray}
Here $m_V$ is the vector-meson mass, the vector-meson decay constant is $f_V= 13.4$
for $V= J/\psi$, and
\begin{eqnarray}
F'_V(t)&=&\frac{1}{m_V^2-t} \left( \frac{2\mu_0^2}{2\mu_0^2 + m_V^2 - t} \right),
\nonumber \\
\label{eq:f1v}
F_1(t) &=& \frac{4m_N^2 - 2.8 t}{(4m_N^2 - t)(1-t/0.71)^2},
\label{eq:f1}
\end{eqnarray}
where $t=(q-k)^2=(p-p')^2$, $t$ is in the unit of GeV$^2$, and $m_N$ is the
nucleon mass.
The propagator $G_{\mathbb P} (s,t)$ of the Pomeron in Eq.~(\ref{eq:MP}) follows the
Regge phenomenology form
\begin{eqnarray}
G_{\mathbb P} = \left(\frac{s}{s_0}\right)^{\alpha_{\mathbb P}(t)-1} \exp
\left\{ - \frac{i\pi}{2} \left[ \alpha_{{\mathbb P}} (t) - 1 \right] \right\},
\label{eq:regge-g}
\end{eqnarray}
where $s=(q+p)^2=W^2$, $\alpha_{\mathbb P} (t) = \alpha_0 + \alpha'_{\mathbb P} t$, and
$s_0=1/\alpha'_{\mathbb P}$.
We use the value of $\alpha' = 0.25$ GeV$^{-2}$ from the works of
Refs.~\cite{Donnachie:1984xq,Donnachie:1992ny,Donnachie:1994zb,Donnachie:1998gm}.
The amplitude $T^{\rm Pom}_{VN,\gamma N}(W)$ has been determined in
Refs.~\cite{Wu:2013xma,Lee:2020iuo,Lee:2022ymp} by fitting the data of total cross
sections up to $W$ = 300 GeV.
The resulting parameters are $\mu_0=  1.1$ GeV$^2$, $\beta_{u/d}=2.07$ GeV$^{-1}$,
$\beta_c = 0.32$ GeV$^{-1}$, and $\alpha_0=1.25$. 
Thus the remaining parameters of the Pom-CQM model in investigating the JLab data are
$\alpha_a$, $\mu_a$, and $N_a$ for $a=$ (B, FSI) of the quark-$N$ potentials given in
Eqs.~(\ref{eq:qn-B})-(\ref{eq:qn-fsi}).

\acknowledgments

The work of S.-H.K. is supported by Basic Science Research Program through the
National Research Foundation of Korea (NRF) under Grants No. 2021R1A6A1A03043957 and
No. 2022R1I1A1A01054390.
T.-S.H.L. and R.B.W. are supported by the Office of Science of the U.S. Department of Energy
under Contract No. DE-AC02-05CH1123.



\begin{thebibliography}{100}

\bibitem{Lee:2022ymp}
T.~S.~H.~Lee, S.~Sakinah, and Y.~Oh,
\newblock Models of $J/\psi$ photo-production reactions on the nucleon,
\href{https://link.springer.com/article/10.1140/epja/s10050-022-00901-9}
{Eur. Phys. J. A \textbf{58}, 252 (2022)}.

\bibitem{GlueX:2019mkq}
A.~Ali \textit{et al.} (GlueX Collaboration),
\newblock First measurement of near-threshold {$J/\psi$} exclusive photoproduction
off the proton,
\href{https://journals.aps.org/prl/abstract/10.1103/PhysRevLett.123.072001}
{Phys. Rev. Lett. \textbf{123}, 072001 (2019)}.

\bibitem{GlueX:2023pev}
S.~Adhikari \textit{et al.} (GlueX Collaboration),
\newblock Measurement of the $J/\psi$ photoproduction cross section over the full
near-threshold kinematic region,
\href{https://journals.aps.org/prc/abstract/10.1103/PhysRevC.108.025201}
{Phys. Rev. C \textbf{108}, 025201 (2023)}.

\bibitem{Duran:2022xag}
B.~Duran \textit{et al.},
\newblock Determining the gluonic gravitational form factors of the proton,
\href{https://www.nature.com/articles/s41586-023-05730-4}
{Nature \textbf{615}, 813 (2023)}.

\bibitem{Brodsky:2000zc}
S.~J.~Brodsky, E.~Chudakov, P.~Hoyer, and J.~M.~Laget,
\newblock Photoproduction of charm near threshold,
\href{https://www.sciencedirect.com/science/article/pii/S0370269300013733?via%3Dihub}
{Phys. Lett. B \textbf{498}, 23 (2001)}.

\bibitem{Lee:2020iuo}
T.~S.~H.~Lee,
\newblock Pomeron-LQCD model of $J/\psi$ photo-production on the nucleon,
\href{https://arxiv.org/abs/2004.13934}
{arXiv:2004.13934 [nucl-th]}.

\bibitem{Mamo:2019mka}
K.~A.~Mamo and I.~Zahed,
\newblock Diffractive photoproduction of $J/\psi$ and $\Upsilon$ using holographic
QCD: Gravitational form factors and GPD of gluons in the proton,
\href{https://journals.aps.org/prd/abstract/10.1103/PhysRevD.101.086003}
{Phys. Rev. D \textbf{101}, 086003 (2020)}.

\bibitem{Du:2020bqj}
M.~L.~Du, V.~Baru, F.~K.~Guo, C.~Hanhart, U.~G.~Mei\ss{}ner, A.~Nefediev, and
I.~Strakovsky,
\newblock Deciphering the mechanism of near-threshold $J/\psi$ photoproduction,
\href{https://link.springer.com/article/10.1140/epjc/s10052-020-08620-5}
{Eur. Phys. J. C \textbf{80}, 1053 (2020)}.

\bibitem{Guo:2021ibg}
Y.~Guo, X.~Ji, and Y.~Liu,
\newblock QCD Analysis of near-threshold photon-proton production of heavy quarkonium,
\href{https://journals.aps.org/prd/abstract/10.1103/PhysRevD.103.096010}
{Phys. Rev. D \textbf{103}, 096010 (2021)}.

\bibitem{JPAC:2023qgg}
D.~Winney \textit{et al.} (Joint Physics Analysis Center),
\newblock Dynamics in near-threshold $J/\psi$ photoproduction,
\href{https://journals.aps.org/prd/abstract/10.1103/PhysRevD.108.054018}
{Phys. Rev. D \textbf{108}, 054018 (2023)}.

\bibitem{Sakinah:2024cza}
S.~Sakinah, T.~S.~H.~Lee, and H.~M.~Choi,
\newblock Dynamical model of $J/\psi$ photoproduction on the nucleon,
\href{https://journals.aps.org/prc/abstract/10.1103/PhysRevC.109.065204}
{Phys. Rev. C \textbf{109}, 065204 (2024)}.

\bibitem{Tang:2024pky}
L.~Tang, Y.~X.~Yang, Z.~F.~Cui, and C.~D.~Roberts,
\newblock $J/\psi$ photoproduction: threshold to very high energy,
\href{https://www.sciencedirect.com/science/article/pii/S0370269324004623?via%3Dihub}
{Phys. Lett. B \textbf{856}, 138904 (2024)}.

\bibitem{Arrington:2021alx}
J.~Arrington \textit{et al.},
\newblock Physics with CEBAF at 12 GeV and future opportunities,
\href{https://www.sciencedirect.com/science/article/pii/S014664102200045X?via%3Dihub}
{Prog. Part. Nucl. Phys. \textbf{127}, 103985 (2022).}

\bibitem{Accardi:2023chb}
A.~Accardi \textit{et al.},
\newblock Strong interaction physics at the luminosity frontier with 22 GeV electrons
at Jefferson Lab,
\href{https://link.springer.com/article/10.1140/epja/s10050-024-01282-x}
{Eur. Phys. J. A \textbf{60}, 173 (2024).}

\bibitem{Liu:2023htv}
T.~Liu, Z.~Zhao, M.~Cai, D.~Byer, and H.~Gao,
\newblock Subthreshold production of $J/\psi$ mesons from the deuteron with the
proposed solenoidal large intensity device,
\href{https://journals.aps.org/prc/abstract/10.1103/PhysRevC.109.065206}
{Phys. Rev. C \textbf{109}, 065206 (2024)}.

\bibitem{Tyson:2024wbx}
R.~Tyson, D.~G.~Ireland, and B.~McKinnon,
\newblock Near-threshold $J/\psi$ photoproduction off the proton and neutron with
CLAS12,
\href{https://www.sif.it/riviste/sif/ncc/econtents/2024/047/04/article/69}
{Nuovo Cim. C \textbf{47}, 212 (2024)}.

\bibitem{Pybus:2024ifi}
J.~R.~Pybus \textit{et al.},
\newblock First measurement of near- and sub-threshold $J/\psi$ photoproduction off
nuclei,
\href{https://arxiv.org/abs/2409.18463}
{arXiv:2409.18463 [nucl-ex]}.

\bibitem{Accardi:2012qut}
A.~Accardi \textit{et al.},
\newblock Electron Ion Collider: The next QCD frontier: Understanding the glue that
binds us all,
\href{https://link.springer.com/article/10.1140/epja/i2016-16268-9}
{Eur. Phys. J. A \textbf{52}, 268 (2016)}.

\bibitem{Abir:2023fpo}
R.~Abir \textit{et al.},
\newblock The case for an EIC theory alliance: Theoretical challenges of the EIC,
\href{https://arxiv.org/abs/2305.14572}
{arXiv:2305.14572 [hep-ph]}.

\bibitem{Wang:2023thy}
X.~Wang, X.~Cao, A.~Guo, L.~Gong, X.~S.~Kang, Y.~T.~Liang, J.~J.~Wu, and Y.~P.~Xie,
\newblock Exclusive charmonium production at the electron-ion collider in China,
\href{https://link.springer.com/article/10.1140/epjc/s10052-024-13033-9}
{Eur. Phys. J. C \textbf{84}, 684 (2024)}.

\bibitem{Kerman:1959fr}
A.~K.~Kerman, H.~McManus, and R.~M.~Thaler,
\newblock The scattering of fast nucleons from nuclei,
\href{https://www.sciencedirect.com/science/article/pii/0003491659900764?via%3Dihub}
{Annals Phys. \textbf{8}, 551 (1959)}.

\bibitem{Landau:1973iv}
R.~H.~Landau, S.~C.~Phatak, and F.~Tabakin,
\newblock Improved theoretical pion-nucleus optical potentials,
\href{https://www.sciencedirect.com/science/article/pii/0003491673902613?via%3Dihub}
{Annals Phys. \textbf{78}, 299 (1973)}.

\bibitem{Lee:1974zr}
T.~S.~H.~Lee and F.~Tabakin,
\newblock Momentum-space study of pion-nucleus inelastic scattering,
\href{https://www.sciencedirect.com/science/article/abs/pii/0375947474904060?via%3Dihub}
{Nucl. Phys. A \textbf{226}, 253 (1974)}.

\bibitem{Feshbach:1992}
H. Feshbach,
\newblock \textit{Theoretical Nuclear Physics: Nuclear Reactions},
{John Wiley and Sons, Inc., New York, 1992}.

\bibitem{Watson:1960}
K.M. Watson,
\newblock Quantum mechanical transport theory. I. Incoherent processes,
\href{https://journals.aps.org/pr/abstract/10.1103/PhysRev.118.886}
{Phys. Rev. \textbf{118}, 886 (1960)}.

\bibitem{Kim:2021adl}
S.~H.~Kim, T.~S.~H.~Lee, S.~i.~Nam, and Y.~Oh,
\newblock Dynamical model of $\phi$ meson photoproduction on the nucleon
and ${^4\rm He}$,
\href{https://journals.aps.org/prc/abstract/10.1103/PhysRevC.104.045202}
{Phys. Rev. C \textbf{104}, 045202 (2021)}.

\bibitem{LEPS:2017nqz}
T.~Hiraiwa \textit{et al.} (LEPS Collaboration),
\newblock First measurement of coherent $\phi$-meson photoproduction from ${^4\rm He}$
near threshold,
\href{https://journals.aps.org/prc/abstract/10.1103/PhysRevC.97.035208}
{Phys. Rev. C \textbf{97}, 035208 (2018)}.

\bibitem{Hatta:2019ocp}
Y.~Hatta, M.~Strikman, J.~Xu, and F.~Yuan,
\newblock Sub-threshold $J/\psi$ and $\Upsilon$ production in $\gamma A$ collisions,
\href{https://www.sciencedirect.com/science/article/pii/S0370269320301258?via%3Dihub}
{Phys. Lett. B \textbf{803}, 135321 (2020)}.

\bibitem{He:2024lry}
F.~He and I.~Zahed,
\newblock Threshold photo-production of $J/\psi$ off light nuclei,
\href{https://arxiv.org/abs/2407.09991}
{arXiv:2407.09991 [nucl-th]}.

\bibitem{Goldberger:1975}
M. L. Goldberger and K. M. Watson,
\newblock \textit{Collision Theory},
{R. E. Krieger Pub. Co., Huntington, New York, 1975}.

\bibitem{Sato:1996gk}
T.~Sato and T.~S.~H.~Lee,
\newblock Meson exchange model for $\pi N$ scattering and $\gamma N \to \pi N$
reaction,
\href{https://journals.aps.org/prc/abstract/10.1103/PhysRevC.54.2660}
{Phys. Rev. C \textbf{54}, 2660 (1996)}.

\bibitem{Matsuyama:2006rp}
A.~Matsuyama, T.~Sato, and T.~S.~H.~Lee,
\newblock Dynamical coupled-channel model of meson production reactions in the nucleon
resonance region,
\href{https://www.sciencedirect.com/science/article/pii/S0370157307000221?via%3Dihub}
{Phys. Rept. \textbf{439}, 193 (2007)}

\bibitem{Kamano:2013iva}
H.~Kamano, S.~X.~Nakamura, T.~S.~H.~Lee, and T.~Sato,
\newblock Nucleon resonances within a dynamical coupled-channels model of $\pi N$ and
$\gamma N$ reactions,
\href{https://journals.aps.org/prc/abstract/10.1103/PhysRevC.88.035209}
{Phys. Rev. C \textbf{88}, 035209 (2013)}.

\bibitem{Segovia:2013wma}
J.~Segovia, D.~R.~Entem, F.~Fernandez, and E.~Hernandez,
\newblock Constituent quark model description of charmonium phenomenology,
\href{https://www.worldscientific.com/doi/abs/10.1142/S0218301313300269}
{Int. J. Mod. Phys. E \textbf{22}, 1330026 (2013)}.

\bibitem{Binkley:1981kv}
M.~E.~Binkley \textit{et al.},
\newblock $J/\psi$ photoproduction from 60 to 300 GeV/c,
\href{https://journals.aps.org/prl/abstract/10.1103/PhysRevLett.48.73}
{Phys. Rev. Lett. \textbf{48}, 73 (1982)}.

\bibitem{E687:1993hlm}
P.~L.~Frabetti \textit{et al.},
\newblock A Measurement of elastic $J/\psi$ photoproduction cross section at Fermilab
E687,
\href{https://www.sciencedirect.com/science/article/abs/pii/037026939390679C?via%3Dihub}
{Phys. Lett. B \textbf{316}, 197 (1993)}.

\bibitem{ZEUS:2002wfj}
S.~Chekanov \textit{et al.} (ZEUS Collaboration),
\newblock Exclusive photoproduction of $J/\psi$ mesons at HERA,
\href{https://link.springer.com/article/10.1007/s10052-002-0953-7}
{Eur. Phys. J. C \textbf{24}, 345 (2002)}.

\bibitem{H1:1996kyo}
S.~Aid \textit{et al.} (H1 Collaboration),
\newblock Elastic and inelastic photoproduction of $J/\psi$ mesons at HERA,
\href{https://www.sciencedirect.com/science/article/pii/055032139600274X?via%3Dihub}
{Nucl. Phys. B \textbf{472}, 3 (1996)}.

\bibitem{H1:2000kis}
C.~Adloff \textit{et al.} (H1 Collaboration),
\newblock Elastic photoproduction of $J/\psi$ and $\Upsilon$ mesons at HERA,
\href{https://www.sciencedirect.com/science/article/pii/S037026930000530X?via%3Dihub}
{Phys. Lett. B \textbf{483}, 23 (2000)}.

\bibitem{Camerini:1975cy}
U.~Camerini, J.~G.~Learned, R.~Prepost, C.~M.~Spencer, D.~E.~Wiser, W.~Ash,
R.~L.~Anderson, D.~Ritson, D.~Sherden, and C.~K.~Sinclair,
\newblock Photoproduction of the $\psi$ Particles,
\href{https://journals.aps.org/prl/abstract/10.1103/PhysRevLett.35.483}
{Phys. Rev. Lett. \textbf{35}, 483 (1975)}.

\bibitem{Wiringa:1994wb}
R.~B.~Wiringa, V.~G.~J.~Stoks, and R.~Schiavilla,
\newblock An accurate nucleon-nucleon potential with charge independence breaking,
\href{https://journals.aps.org/prc/abstract/10.1103/PhysRevC.51.38}
{Phys. Rev. C \textbf{51}, 38 (1995)}.

\bibitem{Piarulli:2016vel}
M.~Piarulli, L.~Girlanda, R.~Schiavilla, A.~Kievsky, A.~Lovato, L.~E.~Marcucci,
S.~C.~Pieper, M.~Viviani, and R.~B.~Wiringa,
\newblock Local chiral potentials with $\Delta$-intermediate states and the structure
of light nuclei,
\href{https://journals.aps.org/prc/abstract/10.1103/PhysRevC.94.054007}
{Phys. Rev. C \textbf{94}, 054007 (2016)}.

\bibitem{Machleidt:1989tm}
R.~Machleidt,
\newblock The meson theory of nuclear forces and nuclear structure,
\href{https://link.springer.com/chapter/10.1007/978-1-4613-9907-0_2}
{Adv. Nucl. Phys. \textbf{19}, 189 (1989)}.

\bibitem{Kamada:2002qt}
H.~Kamada, W.~Gloeckle, J.~Golak, and C.~Elster,
\newblock Lorentz boosted $NN$ potential for few body systems: Application to the
three nucleon bound state,
\href{https://journals.aps.org/prc/abstract/10.1103/PhysRevC.66.044010}
{Phys. Rev. C \textbf{66}, 044010 (2002)}.

\bibitem{Witala:2004pv}
H.~Witala, J.~Golak, W.~Glockle, and H.~Kamada,
\newblock Relativistic effects in neutron-deuteron elastic scattering,
\href{https://journals.aps.org/prc/abstract/10.1103/PhysRevC.71.054001}
{Phys. Rev. C \textbf{71}, 054001 (2005)}.

\bibitem{Witala:2008va}
H.~Witala, J.~Golak, R.~Skibinski, W.~Glockle, W.~N.~Polyzou, and H.~Kamada,
\newblock Relativity and the low-energy $nd$ $A_y$ puzzle,
\href{https://journals.aps.org/prc/abstract/10.1103/PhysRevC.77.034004}
{Phys. Rev. C \textbf{77}, 034004 (2008)}.

\bibitem{Grassi:2022flz}
A.~Grassi, J.~Golak, W.~N.~Polyzou, R.~Skibi\'nski, H.~Wita\l{}a, and H.~Kamada,
\newblock Electron and neutrino scattering off the deuteron in a relativistic
framework,
\href{https://journals.aps.org/prc/abstract/10.1103/PhysRevC.107.024617}
{Phys. Rev. C \textbf{107}, 024617 (2023)}.

\bibitem{Wu:2013xma}
J.~J.~Wu and T.~S.~H.~Lee,
\newblock Production of $J/\psi$ on the nucleon and on deuteron targets,
\href{https://journals.aps.org/prc/abstract/10.1103/PhysRevC.88.015205}
{Phys. Rev. C \textbf{88}, 015205 (2013)}.

\bibitem{Dirac:1949cp}
P.~A.~M.~Dirac,
\newblock Forms of relativistic dynamics,
\href{https://journals.aps.org/rmp/abstract/10.1103/RevModPhys.21.392}
{Rev. Mod. Phys. \textbf{21}, 392 (1949)}.

\bibitem{Keister:1991sb}
B.~D.~Keister and W.~N.~Polyzou,
\newblock Relativistic Hamiltonian dynamics in nuclear and particle physics,
Adv. Nucl. Phys. \textbf{20}, 225 (1991).

\bibitem{Haftel:1970zz}
M.~I.~Haftel and F.~Tabakin,
\newblock Nuclear saturation and the smoothness of nucleon-nucleon potentials,
\href{https://www.sciencedirect.com/science/article/abs/pii/0375947470900473?via%3Dihub}
{Nucl. Phys. A \textbf{158}, 1 (1970)}.

\bibitem{Lonardoni:2017egu}
D.~Lonardoni, A.~Lovato, S.~C.~Pieper, and R.~B.~Wiringa,
\newblock Variational calculation of the ground state of closed-shell nuclei up to
$A=40$,
\href{https://journals.aps.org/prc/abstract/10.1103/PhysRevC.96.024326}
{Phys. Rev. C \textbf{96}, 024326 (2017)}.

\bibitem{Brodsky:1989jd}
S.~J.~Brodsky, I.~A.~Schmidt, and G.~F.~de Teramond,
\newblock Nuclear-bound quarkonium,
\href{https://journals.aps.org/prl/abstract/10.1103/PhysRevLett.64.1011}
{Phys. Rev. Lett. \textbf{64}, 1011 (1990)}.

\bibitem{Gao:2000az}
H.~Gao, T.~S.~H.~Lee, and V.~Marinov,
\newblock $\phi$-$N$ bound state,
\href{https://journals.aps.org/prc/abstract/10.1103/PhysRevC.63.022201}
{Phys. Rev. C \textbf{63}, 022201 (2001)}.

\bibitem{Belyaev:2006vn}
V.~B.~Belyaev, N.~V.~Shevchenko, A.~I.~Fix, and W.~Sandhas,
\newblock Binding of charmonium with two- and three-body nuclei,
\href{https://www.sciencedirect.com/science/article/pii/S0375947406006373?via%3Dihub}
{Nucl. Phys. A \textbf{780}, 100 (2006)}.

\bibitem{Wu:2012wta}
J.~J.~Wu and T.~S.~H.~Lee,
\newblock Photo-production of bound states with hidden charms,
\href{https://doi.org/10.1103/PhysRevC.86.065203}
{Phys. Rev. C \textbf{86}, 065203 (2012)}.

\bibitem{Donnachie:1984xq}
A.~Donnachie and P.~V.~Landshoff,
\newblock Elastic scattering and diffraction dissociation,
\href{https://www.sciencedirect.com/science/article/abs/pii/0550321384903158?via%3Dihub}
{Nucl. Phys. B \textbf{244}, 322 (1984)}.

\bibitem{Donnachie:1992ny}
A.~Donnachie and P.~V.~Landshoff,
\newblock Total cross sections,
\href{https://www.sciencedirect.com/science/article/abs/pii/037026939290832O?via%3Dihub}
{Phys. Lett. B \textbf{296}, 227 (1992)}.

\bibitem{Donnachie:1994zb}
A.~Donnachie and P.~V.~Landshoff,
\newblock Exclusive vector meson production at HERA,
\href{https://www.sciencedirect.com/science/article/pii/0370269395001152?via%3Dihub}
{Phys. Lett. B \textbf{348}, 213 (1995)}.

\bibitem{Donnachie:1998gm}
A.~Donnachie and P.~V.~Landshoff,
\newblock Small $x$: Two pomerons!,
\href{https://www.sciencedirect.com/science/article/pii/S0370269398008995?via%3Dihub}
{Phys. Lett. B \textbf{437}, 408 (1998)}.

\end{thebibliography}
\end{document}